\documentstyle[12pt]{article}

\catcode`\@=11
\def\marginnote#1{}
\newcount\hour
\newcount\minute
\newtoks\amorpm
\hour=\time\divide\hour by60
\minute=\time{\multiply\hour by60 \global\advance\minute
by-\hour}\edef\standardtime{{\ifnum\hour<12
\global\amorpm={am}%
        \else\global\amorpm={pm}\advance\hour by-12 \fi
        \ifnum\hour=0 \hour=12 \fi
        \number\hour:\ifnum\minute<10
0\fi\number\minute\the\amorpm}}
\edef\militarytime{\number\hour:\ifnum\minute<10
0\fi\number\minute}

\def\draftlabel#1{{\@bsphack\if@filesw {\let\thepage\relax
   \xdef\@gtempa{\write\@auxout{\string
      \newlabel{#1}{{\@currentlabel}{\thepage}}}}}\@gtempa
   \if@nobreak \ifvmode\nobreak\fi\fi\fi\@esphack}
        \gdef\@eqnlabel{#1}}
\def\@eqnlabel{}
\def\@vacuum{}
\def\draftmarginnote#1{\marginpar{\raggedright\scriptsize\tt#1}}
\def\draft{\oddsidemargin -.5truein
        \def\@oddfoot{\sl preliminary draft \hfil
        \rm\thepage\hfil\sl\today\quad\militarytime}
        \let\@evenfoot\@oddfoot \overfullrule 3pt
        \let\label=\draftlabel
        \let\marginnote=\draftmarginnote

\def\@eqnnum{(\theequation)\rlap{\kern\marginparsep\tt\@eqnlabel}%
\global\let\@eqnlabel\@vacuum}  }

\def\numberbysection{\@addtoreset{equation}{section}
        \def\theequation{\thesection.\arabic{equation}}}

\def\underline#1{\relax\ifmmode\@@underline#1\else
 $\@@underline{\hbox{#1}}$\relax\fi}

\catcode`@=12
\relax

\numberbysection

\advance \topmargin by -\headheight
\advance \topmargin by -\headsep

\evensidemargin \oddsidemargin
\def\sect#1{\section{#1}}

\def\rf#1{(\ref{#1})}
\def\lab#1{\label{#1}}
\def\nonu{\nonumber}
\def\br{\begin{eqnarray}}
\def\er{\end{eqnarray}}
\def\be{\begin{equation}}
\def\ee{\end{equation}}

\def\lb{\lbrack}
\def\rb{\rbrack}

\def\({\left(}
\def\){\right)}

\newcommand{\ct}[1]{\cite{#1}}
\newcommand{\bi}[1]{\bibitem{#1}}

\relax

\newcommand\sbr[2]{\left\lbrack\,{#1}\, ,\,{#2}\,\right\rbrack}

\def\a{\alpha}

\def\b{\beta}

\def\d{\delta}

\def\eps{\epsilon}
\def\bareps{{\bar \epsilon}}

\def\h{{1\over 2}}
\def\l{\lambda}

\def\o{\over}

\def\pa{\partial}
\def\pr{\prime}

\def\ra{\rightarrow}

\def\tp0{\Theta_{+}^{(0)}}
\def\tm0{\Theta_{-}^{(0)}}

\def\vp{\varphi}

\def\cgh{{\hat {\cal G}}}
\def\ch{{\cal H}}

\def\cl{{\cal L}}

\def\f#1#2#3 {f^{#1#2}_{#3}}

\def\win1{{\sf w_{1+\infty}}}

\def\Win1{{\sf W_{1+\infty}}}

\def\rlx{\relax\leavevmode}
\def\inbar{\vrule height1.5ex width.4pt depth0pt}
\def\IZ{\rlx\hbox{\sf Z\kern-.4em Z}}
\def\IR{\rlx\hbox{\rm I\kern-.18em R}}
\def\IC{\rlx\hbox{\,$\inbar\kern-.3em{\rm C}$}}
\def\IN{\rlx\hbox{\rm I\kern-.18em N}}
\def\IO{\rlx\hbox{\,$\inbar\kern-.3em{\rm O}$}}
\def\IP{\rlx\hbox{\rm I\kern-.18em P}}
\def\IQ{\rlx\hbox{\,$\inbar\kern-.3em{\rm Q}$}}
\def\IF{\rlx\hbox{\rm I\kern-.18em F}}
\def\IG{\rlx\hbox{\,$\inbar\kern-.3em{\rm G}$}}
\def\IH{\rlx\hbox{\rm I\kern-.18em H}}
\def\II{\rlx\hbox{\rm I\kern-.18em I}}
\def\IK{\rlx\hbox{\rm I\kern-.18em K}}
\def\IL{\rlx\hbox{\rm I\kern-.18em L}}
\def\one{\hbox{{1}\kern-.25em\hbox{l}}}
\def\0#1{\relax\ifmmode\mathaccent"7017{#1}%
B        \else\accent23#1\relax\fi}

\def\NPB#1#2#3{{\sl Nucl. Phys.} {\bf B#1} (#2) #3}
\def\NPBFS#1#2#3#4{{\sl Nucl. Phys.} {\bf B#2} [FS#1] (#3) #4}

\def\PRD#1#2#3{{\sl Phys. Rev.} {\bf D#1} (#2) #3}

\def\PLB#1#2#3{{\sl Phys. Lett.} {\bf #1B} (#2) #3}
\def\JMP#1#2#3{{\sl J. Math. Phys.} {\bf #1} (#2) #3}

\def\IJMPA#1#2#3{{\sl Int. J. Mod. Phys.} {\bf A#1} (#2) #3}

\begin{document}
\begin{titlepage}
\vspace*{-1cm}

\noindent
September, 1999 \hfill{IFT-P.073/99}\\
 \hfill{hep-th/9909118}  

\vskip 3cm

\vspace{.2in}
\begin{center}
{\large\bf Confinement, solitons  and the equivalence between \\
the sine-Gordon and massive Thirring models}
\end{center}

\vspace{1in}

\begin{center}
H.S. Blas Achic and L. A. Ferreira

\vspace{.5 cm}
\small

\par \vskip .1in \noindent
Instituto de F\'\i sica Te\'orica - IFT/UNESP\\
Rua Pamplona 145\\
01405-900  S\~ao Paulo-SP, BRAZIL

\normalsize
\end{center}

\vspace{1.5in}

\begin{abstract}
We consider a two-dimensional integrable and conformally invariant field
theory possessing two Dirac spinors and three scalar fields. The interaction
couples bilinear terms in the spinors to exponentials of the scalars. Its
integrability properties are based on the $sl(2)$ affine Kac-Moody algebra,
and it is a simple example of the so-called conformal affine Toda theories
coupled to matter fields.  We show, using bosonization techniques,  that 
the classical equivalence
between a $U(1)$ Noether current and the topological current holds true at the
quantum level, and then leads to a bag model like mechanism for the
confinement of the spinor fields inside the solitons. By bosonizing the
spinors we show that the theory decouples into a  sine-Gordon model and free
scalars. We construct the two-soliton solutions and show that their
interactions lead to the same time delays as those for the sine-Gordon
solitons. The model provides a good laboratory to test duality ideas in the
context of the equivalence between the sine-Gordon and Thirring theories. 

\end{abstract}
\end{titlepage}

\sect{Introduction}

Solitons are believed to have an important role in many non-perturbative
aspects of a wide class of quantum field theories, as well as in condensed
matter phenomena. The interest is greater in Lorentz invariant theories
presenting topological solitons, since in many cases there are strong evidence
that such solitons correspond to particle excitations in the quantum spectrum
of the theory. The relevance of the solitons to non-perturbative phenomena
comes, in general, from the fact that their interactions are inversely
proportional to the  coupling constants governing the dynamics of the
fundamental fields appearing in the Lagrangean. Therefore, the solitons are
weakly coupled in the strong regime of the theory, and that is the basic fact
underlying several duality ideas. The reason is that one can describe the
theory in the strong coupling regime, by replacing the fundamental Lagrangean
by another one where the excitations of its fields correspond to the solitonic
states. The electromagnetic duality of Montonen-Olive \ct{mo}, involving
magnetic monopoles and gauge particles, is the best
example of 
such behaviour, and has found in supersymmetric gauge theories the best habitat
for its implementation \ct{vw,sw,sen}.  Similar dualities occur in
statistical mechanics models and many lower dimensional field theories. The
example which is best understood in such context is the quantum equivalence
between the sine-Gordon and Thirring models \ct{coleman,mandelstam}, which
provide an excellent laboratory to test ideas about the role of solitons in
quantum field theories. 

In this paper we consider a two dimensional integrable and conformally
invariant field theory involving spinor and scalar fields, and presenting
topological solitons solutions. The theory is one of the simplest examples of
the so-called Conformal Affine Toda models coupled to matter fields  
proposed in \ct{matter}.  It has two Dirac spinors coupled to a scalar field
which plays the role of a $U(1)$ ``gauge'' particle. The system is made
conformally invariant by the introduction of two other scalar fields
constituting what is in general called in the literature a beta-gamma system,
with non-positive definite kinetic term. The integrability of the theory is
established using a zero curvature formulation of its equations of motion
based on the ${\widehat {sl}}(2)$ affine Kac-Moody algebra. The general
solution as well as  
explicit one-soliton solutions have been obtained in \ct{matter}. Besides the
conformal and local gauge symmetries, the model presents chiral symmetry and
some discrete symmetries. However, one of its main properties is that for a
large subset of solutions there is an equivalence between the $U(1)$ Noether
gauge current, involving the spinors only, and the topological current
associated to the scalar ``gauge'' particle. That fact was stablished in
\ct{matter} at the classical level, and has profound consequences in the
properties of the theory. It implies that the density of the $U(1)$ charge has
to be concentrated in the regions where the scalar field has non vanishing
space derivative, i.e. presents large momenta modes. The one-soliton of the
theory is of the sine-Gordon type, and therefore the charge density is
concentrated inside the soliton. Since the charge carriers are the spinors, it
means that if one looks for excitations of the theory around the solitonic
state, one would expect the spinors to be confined. 
 
One of the main objectives of this paper is to study the equivalence of the
Noether and topological currents at the quantum level, and its consequences
for the confinement mechanism. The first thing to be established is if such
equivalence is not spoiled by quantum anomalies the currents may
present. Fortunately, that issue can be established exactly by using, instead
of perturbative approaches, bosonization techniques following the lines of
\ct{witten}. By bosonizing the spinor field we show that the sector of the
theory made of the spinor and ``gauge'' particles is equivalent to a theory of
a free massless scalar and a sine-Gordon field. In addition, the condition for
the equivalence of the Noether and topological currents is simply the
condition that the free massless scalar should be constant.  Therefore,
by performing a quantum reduction where the excitations of the free scalar are
eliminated, we obtain a submodel where the equivalence of those currents holds
true exactly at the quantum level, proving that there are no
anomalies. Consequently, we show that in such reduced theory the confinement
of the spinor particles does take place. 

An important property of the model under consideration is that it possesses
another type of spinor particle. That is obtained by fermionizing  the
sine-Gordon field using the well known equivalence between the sine-Gordon and
massive Thirring models \ct{coleman,mandelstam}. In that scenario the solitons
of the sine-Gordon model are interpreted as the spinor particles of the
Thirring theory. We are then lead to an interesting analogy with what one
expects to happen in QCD. The original spinor particles of our model that get
confined inside the solitons play the
role of the quarks, and the second spinor particle (Thirring), which are the
solitons, play the role of the hadrons. The $U(1)$ Noether charge is also
confined and is analogous to color in QCD. In this sense, our model constitute
an excellent laboratory to test ideas about confinement, the role of solitons
in quantum field theory, and dualities interchanging the role of solitons and
fundamental particles.   

We also constructed the one-soliton and two-soliton solutions for our theory
using  two techniques: the dressing transformations and the Hirota's
method. We use the zero curvature representation of the equations of motion
with potentials taking value on the affine Kac-Moody algebra 
$\widehat{sl} (2)$ (see appendix). Then the basic idea is to look for vacuum
configurations where the potentials lie in an abelian (up to central terms) 
subalgebra of $\widehat{sl} (2)$. That constitutes in fact an algebra of
oscillators. The solitons are obtained by performing the
dressing tranformations from those vacuum configurations with group elements
which are exponentiations of the 
eigenvectors of the oscillators \ct{tau,otu,ous}. Such procedure leads quite
naturally to the definition of tau-functions \ct{tau}, and then the Hirota's
method can be easily implemented too. We discuss the conditions for the
solutions to be real, and evaluate the topological charges. The interactions of
the solitons is studied by calculating their time delays. It is
attractive, in fact the time delays are the same as those for the
sine-Gordon solitons. An interesting aspect of the  solutions is that when the
two Dirac spinors of the theory are  related by a reality
condition, then either the soliton or the anti-soliton disappears from the
spectrum.  We interpret that as  indicative of the existence of a duality
involving the solitons and the spinor particles. 

The paper is organized as follows: in section \ref{sec:model} we summarize the
properties of 
the model, introduced in \ct{matter}, at the classical level. In section
\ref{sec:quantum} we consider the correponding  quantum field theory  and
show, using bosonization techniques,  the equivalence of the Noether and
topological currents as well as the confinement mechanism. The soliton
solutions and their interaction (time delays) are studied in section
\ref{sec:soliton}. In the appendix we give the zero curvature representation
for the model under consideration.

\sect{The model}
\label{sec:model}

Consider the two-dimensional field theory defined by the Lagrangian
\ct{matter} 
\br
{1\o k}\, \cl = - {1\o 4} \pa_{\mu} \vp \, \pa^{\mu} \vp
+ \h  \pa_{\mu} \nu \, \pa^{\mu} \eta
- {1\o 8}\, m_{\psi}^2 \, e^{2\,\eta} 
+ i  {\bar{\psi}} \gamma^{\mu} \pa_{\mu} \psi
- m_{\psi}\,  {\bar{\psi}} \,
e^{\eta+2i\vp\,\gamma_5}\, \psi
\lab{lagrangian}
\er
where $\vp$, $\eta$ and $\nu$ are scalar fields, and $\psi$ is a
Dirac spinor. Notice that, ${\bar{\psi}} \equiv {\widetilde \psi}^{T} \,
\gamma_0$, with ${\widetilde \psi}$ being a second Dirac spinor. However, in
many applications in this paper we shall take  ${\widetilde \psi}$  to satisfy
the reality condition
\be
{\widetilde \psi} = e_{\psi} \,\psi^*
\lab{psitrealcond}
\ee
where $e_{\psi}$ is a real dimensionless  constant. 
As we will show, the sign of
$e_{\psi}$ will have  an important role in  determining the spectrum of
soliton solutions.  The corresponding equations of motion are
\br
\pa^2 \,\vp &=& i4 m_{\psi}\,  \overline
\psi \gamma_5 e^{\eta+2i\varphi \gamma_5} \psi,
\lab{sl2eqm1}\\
\pa^2\,{\nu} &=& -  2 m_{\psi}\,  \overline
\psi e^{\eta+2i\varphi \gamma_5} \psi
 - {1\over 2}  m_{\psi}^2 e^{2\eta},
\lab{sl2eqm2}\\
\pa^2 \eta &=& 0,
\lab{sl2eqm3}\\
i \gamma^{\mu} \pa_{\mu} \psi &=& m_{\psi}\, e^{\eta+2i\vp\,\gamma_5} \,
\psi ,\lab{sl2eqm4}\\
i \gamma^{\mu} \pa_{\mu} \widetilde \psi &
=& m_{\psi}\, e^{\eta-2i\vp\,\gamma_5} \,
\widetilde  \psi ,
 \lab{sl2eqm5}
\er

The theory \rf{lagrangian} was proposed in \ct{matter} as an example of a wide
class of integrable theories called Affine Toda systems coupled to matter
fields\footnote{The Lagrangian \rf{lagrangian} is obtained from (10.18) of 
\ct{matter} by the replacements $\vp \ra i \vp$ and ${\tilde{\nu}}\ra \nu -
{i\o 2} \vp$.}. 
The zero curvature representation, the construction of the general
solution including the solitonic ones and many other properties were discussed
in \ct{matter}. In the appendix \ref{appa} we summarize some of those
results. Here, we want to discuss some special features of that theory at the
quantum level as well as the two-soliton solutions. We start by reviewing the
symmetries of \rf{lagrangian}. 

{\em Conformal symmetry}. The model  \rf{lagrangian} is invariant under the
conformal transformations\footnote{We are using $x_{\pm}=t \pm x$, and so,
$\pa_{\pm} = \h \( \pa_t \pm \pa_x\)$, and 
$\pa^2 = \pa_t^2 - \pa_x^2 = 4 \pa_{+}\pa_{-}$.}
\be
x_{+} \ra {\hat x}_{+} = f(x_{+}) \, , \qquad 
x_{-} \ra {\hat x}_{-} = g(x_{-}),
\lab{ct}
\ee
with $f$ and $g$ being analytic functions; and with the fields transforming 
as\footnote{We take 
$\gamma_0 = -i \(
\begin{array}{rr} 0&-1\\ 1&0
\end{array}\)$ , 
$\gamma_1 = -i \(
\begin{array}{rr}  0&1\\  1&0
\end{array}\)$, 
$\gamma_5 = \gamma_0\gamma_1 = \(
\begin{array}{rr} 1&0\\ 0&-1
\end{array}\)$}
\br
\vp (x_{+}\, , \, x_{-}) &\ra& 
{\hat {\vp}}({\hat x}_{+}\, , \,  {\hat x}_{-}) = 
\vp (x_{+}\, , \, x_{-}) \, ,
\nonu\\
e^{-\nu (x_{+}\, , \, x_{-})} &\ra& 
e^{-{\hat \nu}({\hat x}_{+}\, , \, 
{\hat x}_{-})} = \( f^{\pr}\)^{\d} \, \( g^{\pr}\)^{\d}
e^{-\nu (x_{+}\, , \, x_{-})} \, ,
\lab{ctf}\\
e^{-\eta (x_{+}\, , \, x_{-})} &\ra& e^{-{\hat \eta}({\hat x}_{+}\, , \, 
{\hat x}_{-})} = \( f^{\pr}\)^{\h} \, \( g^{\pr}\)^{\h}  e^{-\eta (x_{+}\, 
, \, x_{-})} \, ,
\nonu\\ 
\psi (x_{+}\, , \, x_{-}) &\ra & {\hat {\psi}} ({\hat x}_{+}\, , \, 
{\hat x}_{-}) =   e^{{1\o 2}\( 1+ \gamma_5\) \log \( f^{\pr}\)^{-\h} 
+ {1\o 2}\( 1- \gamma_5\) \log \( g^{\pr}\)^{-\h}}
\, \psi (x_{+}\, , \, x_{-}) \, ,
\nonu 
\er
where the conformal weight $\d$, associated to $e^{-\nu}$, is arbitrary, and 
$\widetilde \psi$ transforms in the same way as $\psi$.

{\em Left-right local symmetries}. The Lagrangian \rf{lagrangian} is invariant
under the local $U(1)_L \otimes U(1)_R$ transformations
\be
\vp \ra \vp + \xi_{+}\( x_{+}\) + \xi_{-}\( x_{-}\) \; ; \qquad 
\nu \ra \nu \; ; \qquad \eta \ra \eta 
\ee
and
\be 
\psi \ra e^{- i\( 1+ \gamma_5\) \xi_{+}\( x_{+}\) 
+ i\( 1- \gamma_5\) \xi_{-}\( x_{-}\)}\, \psi 
\; ; \qquad 
{\widetilde \psi} \ra e^{ i\( 1+ \gamma_5\) \xi_{+}\( x_{+}\) 
- i\( 1- \gamma_5\) \xi_{-}\( x_{-}\)} {\widetilde \psi} 
\ee

{\em $U(1)$ global symmetry}. Notice that, by taking $\xi_{+}\( x_{+}\) = -
\xi_{-}\( x_{-}\) = - \h 
\theta$, with $\theta = {\rm const.}$, one gets a global $U(1)$ transformation 
\be
\vp \ra \vp  \; ; \qquad 
\nu \ra \nu \; ; \qquad \eta \ra \eta   \; ; \qquad 
\psi \ra e^{i \theta} \, \psi  \; ; \qquad 
{\widetilde \psi} \ra e^{-i \theta} \, {\widetilde \psi} 
\lab{globalu1}
\ee
The corresponding Noether current is given by
\be
J^{\mu} = {\bar{\psi}}\, \gamma^{\mu}\, \psi \, , \qquad
\pa_{\mu}\, J^{\mu} = 0.
\lab{noethersl2}
\ee

{\em Chiral symmetry}. In addition, if one takes $\xi_{+}\( x_{+}\) = 
\xi_{-}\( x_{-}\) = - \h \a$,  with $\a = {\rm const.}$, one gets  the global
chiral symmetry 
\be
\psi \ra e^{i\gamma_5 \a}\, \psi \; ; \qquad 
{\widetilde \psi}\ra e^{-i\gamma_5 \a}\, {\widetilde \psi} \; ; \qquad 
\vp \ra \vp -  \a \; ; \qquad 
\nu \ra \nu \; ; \qquad \eta \ra \eta 
\lab{chiraltransf}
\ee
with the corresponding Noether current 
\be
J_5^{\mu} =  \bar \psi \gamma_5 \gamma^\mu \psi 
+{1\over 2} \partial^\mu \vp  
\; ; \qquad \qquad \pa_{\mu}J_5^{\mu} =0 
\lab{chiral}
\ee

{\em Topological charge}. One can shift the $\vp$ field as $\vp \ra \vp + n
\pi$, keeping all the other fields unchanged, that the Lagrangian is left 
invariant. That means that the  theory possesses an infinite number of vacua, 
and the topological charge 
\be
Q_{\rm topol.} \equiv \int \, dx \, j^0 
\, , \qquad
j^{\mu} =  {1\o{\pi }}\epsilon^{\mu\nu} \pa_{\nu} \, \vp 
\lab{topological}
\ee
can assume non trivial values. 

{\em  CP-like symmetry}. Finally the Lagrangian \rf{lagrangian} is invariant
under the transformation
\be
x_{+} \leftrightarrow x_{-} \, ; \quad \psi 
\leftrightarrow i\epsilon  \gamma_0{\widetilde {\psi}} \, ; \quad  
{\widetilde {\psi}} \leftrightarrow -i\epsilon   \gamma_0 \psi \, ; \quad 
\vp \leftrightarrow \vp \, ; \quad 
\eta \leftrightarrow \eta \, ; \quad 
\nu \leftrightarrow \nu 
\lab{cpsymmetry}
\ee
where $\epsilon = \pm 1$. Notice that by imposing the reality condition
\rf{psitrealcond} one breaks such CP symmetry, for any real value of the
constant $e_{\psi}$\footnote{Notice that the CP symmetry is preserved if one
takes $e_{\psi}=\pm i$.}. 
 
Now comes a very interesting property of this model. The conservation of the
$U(1)$ vector current \rf{noethersl2} and of the chiral current \rf{chiral}
can be used to show that there exist two charges given by 
\be
{\cal J}= - {\widetilde {\psi}}^T \( 1+\gamma_5 \)\psi 
+ \partial_+\vp, \qquad
{\bar {\cal J}}=  {\widetilde {\psi}}^T \( 1-\gamma_5 \)\psi 
+ \partial_-\vp 
\lab{chiralcur}
\ee
satisfying 
\be
\partial_-{\cal J}=0 \; ; \qquad \quad  \partial_+{\bar {\cal J}}=0
\ee
Notice, from \rf{ctf}, that the currents ${\cal J}$ and ${\bar {\cal J}}$ have
conformal weights $(1,0)$ and $(0,1)$ respectively. 
One can now perform a (Hamiltonian) reduction of the model by imposing the
constraints 
\be
{\cal J}=0 \; ; \qquad \quad {\bar {\cal J}}=0 
\lab{constraints}
\ee
The degree of freedom eliminated by such reduction does not correspond to the
excitations of any of the fields appearing in the Lagragian
\rf{lagrangian}. As we show below, it corresponds to the excitations of a free
field which is a non-linear combination of those in \rf{lagrangian}. 

One can easily check that the constraints \rf{constraints} are equivalent to 
\be
{1\o{2\pi }}\epsilon^{\mu\nu} \pa_{\nu} \, \vp=
{1\o \pi} \bar \psi \gamma^\mu  \psi,
\lab{equivcurrents}
\ee
Therefore, in the reduced model, the Noether  current \rf{noethersl2} is
proportional to the topological current \rf{topological}. 

That fact has profound consequences in the properties of such theory. For
instance, it implies (taking ${\widetilde {\psi}}$ to be the complex
conjugate of $\psi$) that the charge density $\psi^{\dagger}\psi$ is
proportional to the space derivative of $\vp$. Consequently, the Dirac field
is confined to live in regions where the field $\vp$ is not constant. The best
example of that is the 
one-soliton solution of \rf{lagrangian} which was calculated in \ct{matter}
and it is given by
\br
\vp &=& 2 \arctan \( \exp \( 2m_\psi \( x-x_0-vt\)/\sqrt{1-v^2}\)\) \nonu\\
\psi &=& e^{i\theta} \sqrt{m_\psi} \, 
e^{m_\psi \( x-x_0-vt\)/\sqrt{1-v^2}}\, \(
\begin{array}{c}
\left( { 1-v\o 1+v}\right)^{1/4}  
{1 \o 1 + ie^{2 m_\psi \( x-x_0-vt\)/\sqrt{1-v^2}}}\\
-\left( { 1+v\o 1-v}\right)^{1/4}  
{1 \o 1 - ie^{2 m_\psi \( x-x_0-vt\)/\sqrt{1-v^2}}} 
\end{array}\) 
\nonu\\
\nu &=& 
- \h \log \( 1 + \exp \( 4 m_\psi \( x-x_0-vt\)/\sqrt{1-v^2}\)\) 
- {1\o 8} m_{\psi}^2 x_{+} x_{-}
\nonu\\
\eta & = & 0 
\lab{solsimple}
\er
and the solution for $\widetilde \psi$ is the complex conjugate of
$\psi$. Notice that, from \rf{topological}, one indeed has $Q_{\rm topol.} =1$
for the  solution \rf{solsimple}\footnote{Notice that we have defined the
topological charge in \rf{topological} as twice that of the reference
\ct{matter}, in order to make it integer.}.  
 
Notice that the solution for $\vp$ is exactly the sine-Gordon soliton, and
therefore $\pa_x \vp$ is non-vanishing only in a region of size of the order
of $m_\psi^{-1}$. In addition, the solution for $\psi$ satisfies the massive
Thirring model equations of motion \ct{orfanidis}. 
One can check that \rf{solsimple} satisfies
\rf{equivcurrents}, and so is a solution of the reduced model. Therefore, the
Dirac field must be confined inside the soliton. One can verify that 
\rf{solsimple}  indeed confirms that fact. 

We point out that the condition \rf{equivcurrents} together with the equations
of motion for the Dirac spinors \rf{sl2eqm4}-\rf{sl2eqm5} imply the equation
of motion for $\vp$, namely \rf{sl2eqm1}. Therefore in the reduced model,
defined by the constraints \rf{constraints}, one can replace a second order
differential equation, i.e. \rf{sl2eqm1}, by two first order equations,
i.e. \rf{equivcurrents}. 

One could think of equating the chiral current \rf{chiral} to a topological
current associated to a scalar field. However, the equations of motion imply
that such a scalar has to be a free field. Therefore, one could consider a
submodel of \rf{lagrangian} defined by the constraints 
\be
\l \, \epsilon^{\mu\nu} \pa_{\nu} \, \eta =   
\bar \psi \gamma_5 \gamma^\mu \psi +{1\over 2} \partial^\mu \vp
\ee
where $\l$ is a parameter. One can check that such constraints are compatible
with the equations of motion. However, we will not use that observation in
what follows. 

\sect{The quantum theory}
\label{sec:quantum}

The results discussed above are true at the classical level.  
The question we address now is if they remain true at the quantum level, an
the 
confinement of the Dirac spinor does take place. The answer is affirmative, and
can be obtained quite elegantly using bosonization methods
\ct{coleman,mandelstam}. Witten has 
considered a similar model in \ct{witten}, originally proposed by Kogut and
Sinclair \ct{kogut}, and we shall use his methods. Their 
model differs from \rf{lagrangian} in three points: {\em i)} it does not
contain 
the pair of fields $\( \eta , \nu\)$, and therefore is not
conformally invariant; {\em ii)} the sign of the kinetic term of the $\vp$
field 
in \rf{lagrangian} is flipped with respect to their corresponding term;  
{\em iii)} their model possesses just one Dirac spinor. 

In the considerations  about the quantum theory, to be made in this
section, we shall impose the reality condition \rf{psitrealcond} and
consequently the Lagrangian  \rf{lagrangian} will depend on the real constant
$e_{\psi}$. Therefore, in this section we have ${\bar{\psi }}= \psi^{\dagger}
\gamma_0$. 

We introduce a new  boson field $c$ by
bosonizing the Dirac spinor $\psi$ as \ct{witten,abdalla}  
\br
i: {\bar{\psi }}\gamma ^{\mu }\partial_{\mu }\psi : &=&
\frac{\a^2}{2\pi} : (\partial_{\mu }c)^{2} : 
\lab{boson1}\\
: {\bar{\psi }}(1\pm \gamma _{5})\psi : &=& 
- {\mu \o 2 \pi} :  e^{(\pm i 2 \a c)} :  
\lab{boson2}\\
: {\bar{\psi }}\gamma ^{\mu }\psi : &=&-\frac{\a}{\pi }
\epsilon^{\mu \nu}\partial _{\nu }c  
\lab{boson3}
\er
where $\a$ is a real parameter, and $\mu$ is a mass parameter used as an
infrared regulator. 

Rewriting the last term of \rf{lagrangian} as
\be
{\bar{\psi}} \, e^{\eta+2i\vp\,\gamma_5}\, \psi = 
e^{\eta}\, \( {\bar{\psi}}  {\( 1 + \gamma_5\)\o 2} \psi \, e^{2i\vp} + 
{\bar{\psi}} {\( 1 - \gamma_5\)\o 2} \psi \, e^{-2i\vp} \) 
\ee
one then obtains that the Lagrangian \rf{lagrangian} becomes
\br
{1\o k}\,  \cl &=& - {1\o 4} \( \pa_{\mu} \vp \)^2 
+ \h  \pa_{\mu} \nu \, \pa^{\mu} \eta
- {m_{\psi}^2 \o 8}\, e^{2\,\eta} \nonu\\ 
&+& e_{\psi}\( \frac{\a^2}{2\pi}(\partial_{\mu }c)^{2}
+  {\mu  m_{\psi} \o 4 \pi} \, e^{\eta}\, \( e^{2i\( \a c + \vp\)} + 
e^{-2i\( \a c + \vp\)}\) \) 
\lab{lagraboson1}
\er 

Introduce the linear combinations
\br
\phi \equiv 2 \,\,  {\( \a\, c +  \vp \) \o {\sqrt{\mid 4 \pi - 
8 e_{\psi}\mid}}} 
\; ; \qquad \qquad 
\rho \equiv \sqrt{2\o \pi}\, 
{\( 2 e_{\psi}\a \, c + \pi \vp\)  \o {\sqrt{\mid 4 \pi - 8e_{\psi}\mid}}} 
\lab{redeffields}
\er
After rescaling the fields as 
\be
\rho \ra {\rho\o \sqrt{k}} \qquad \qquad \nu \ra {\nu \o k} \qquad \qquad 
\phi \ra {\phi \o \sqrt{k}}
\ee 
we rewrite the Lagrangian as 
\br
 \cl = -  {1\o 2}\, \epsilon\( e_{\psi}\) \( \pa_{\mu} \rho \)^2 
+ \h  \pa_{\mu} \nu \, \pa^{\mu} \eta
- {k\o 8}\, m_{\psi}^2 \, e^{2\,\eta} 
+  
e_{\psi}\( \frac{1}{2}\, \epsilon\( e_{\psi}\) \(\partial_{\mu }\phi \)^{2}
+  {m^2 \o \b^2}\, e^{\eta}\,  \cos \(  \b \, \phi \) \) 
\lab{lagraboson2}
\er 
where we have introduced
\be
\b \equiv \sqrt{\mid 4 \pi - 8 e_{\psi}\mid \o k}\; ; \qquad 
m^2 \equiv {\mu \o 2 \pi} \mid 4 \pi - 8 e_{\psi} \mid  m_{\psi}  
\; ; \qquad  
\epsilon\( e_{\psi}\) \equiv {\rm sign} \, \( 4 \pi - 8 e_{\psi}\) 
\lab{betadef}
\ee
Therefore, $\rho$ is a free scalar field and $\phi$ is a sine-Gordon
field coupled to $\eta$. However, in order for the kinetic and potential terms 
for $\phi$ to lead to a unitary sine-Gordon theory  we shall impose 
\be
\epsilon\( e_{\psi}\) = 1 \; ; \qquad \qquad e_{\psi} < \frac{\pi}{2}
\lab{condepsi}
\ee
Clearly, in the limit  $e_{\psi} \ra \pi /2$, $\phi$ becomes a massless free
field.

We now discuss the quantum version of the reduction \rf{constraints}.  
Since ${\widetilde {\psi}}^T \( 1\pm \gamma_5 \)\psi = e_{\psi} 
{\bar \psi} \( \gamma_0 \pm \gamma_1 \) \psi$ (see \rf{psitrealcond}), 
it follows from \rf{chiralcur}, 
\rf{boson3} and \rf{redeffields} that ($\epsilon^{01} = 1$) 
\be
{\cal J} = {\sqrt{\mid 4 \pi - 8 e_{\psi}\mid }\o \sqrt{2 \pi}} 
\, \pa_{+} \rho  
\; ; \qquad \qquad 
{\bar {\cal J}}= {\sqrt{\mid 4 \pi - 8 e_{\psi}\mid }\o \sqrt{2 \pi}} 
\, \pa_{-} \rho
\ee
Therefore, the constraints \rf{constraints} are equivalent to 
$\rho = {\rm const.}$, and consequently the degree of freedom elimaneted by
them corresponds to the $\rho$ field. 
  
The reduction at the quantum level can be realized as follows. Since the
scalar field $\rho$ is decoupled from all others fields in the theory
\rf{lagraboson2}, we can denote the  space of states as 
$\ch = \ch_{\rho}\otimes \ch_0$, 
where $\ch_{\rho}$ is the Fock space of the free massless scalar
$\rho$, and $\ch_0$ carries the states of the rest of the theory. We shall
denote $\rho = \rho^{(+)} + \rho^{(-)}$, where $\rho^{(+)}$ ($\rho^{(-)}$)
corresponds to the 
part containing annihilation (creation) operators in its expansion on plane
waves.   The
reduction corresponds to the  restriction of the theory to those states
satisfying  
\be
\pa_{\pm} \rho^{(+)} \mid \Psi \rangle = 0 
\lab{reduce}
\ee
By taking the complex conjugate one gets
\be
\langle \Psi \mid  \pa_{\pm} \rho^{(-)} = 0
\lab{reducea}
\ee
and consequently the expectation value of $\pa_{\pm} \rho $ vanishes on such
states. Indeed, if $\mid \Psi \rangle $ and $\mid \Psi^{\pr} \rangle $ are two
states satisfying \rf{reduce}, then 
\be
\langle \Psi^{\pr} \mid \pa_{\pm} \rho \mid \Psi \rangle  = 
\langle \Psi^{\pr} \mid \pa_{\pm} \rho^{(-)} + \pa_{\pm} \rho^{(+)} \mid \Psi
\rangle  
= 0 
\lab{reduceb}
\ee
That provides the correspondence with the classical constraints
\rf{constraints}. Therefore,  the Hilbert space of the reduced theory is
$\ch_{c} = \mid \Psi  \rangle \otimes \ch_0$. 

For the theory described by $\ch_{c}$, the equivalence between Noether and
topological currents, given by \rf{equivcurrents}, holds true at the quantum
level, since as we have shown before, \rf{equivcurrents} is equivalent to the
vanishing of the currents ${\cal J}$ and ${\bar {\cal J}}$. Notice that such
quantum equivalence is exact, since we have not used 
perturbative or semiclassical methods.  

One of the consequences of that quantum equivalence is that in the states of 
$\ch_{c}$, like the one-soliton \rf{solsimple}, where the the space derivative
of $\vp$ is localized\footnote{Notice that for the states of $\ch_{c}$, where
\rf{reduceb} holds true, one has from \rf{redeffields} that $\langle \pa_{\pm}
\phi \rangle \sim \langle \pa_{\pm} \vp \rangle$.}, one has that the spinor
$\psi$ is confined, since from \rf{equivcurrents}, $\langle \pa_x \vp \rangle
\sim \langle \psi^{\dagger} \psi \rangle$. Therefore, we have shown that the
confinement of $\psi$ does take place in the quantum  theory. 

Notice that the quantum reduction \rf{reduce} does not violate the conformal
symmetry of the Lagrangian \rf{lagraboson2}, because we are not restricting
to a non vanishing constant, the expectation value of any quantity of non
vanishing conformal weight. 

The theory described by \rf{lagraboson2} is certainly non-unitary because the
kinetic terms have in general different signs, and so the energy is unbounded
from below\footnote{For the choice $e_{\psi} <0$, the kinetic terms for $\rho$
and $\phi$ have the same sign.}. The reduction \rf{reduce} eliminates part of
the problem, since 
the kinetic energy associated to the $\rho$ field vanishes in $\ch_{c}$. There
remains the non-unitarity associated with the system $\( \eta , \nu\)$. We
can perform a further reduction by restricting ourselves to those states of
$\ch_0$ satisfying
\be
\pa_{\pm} \eta^{(+)} \mid \Psi_0 \rangle = 0 
\lab{reduceeta}
\ee 
where  we have denoted the free field $\eta$ as $\eta = \eta^{(+)} +
\eta^{(-)}$, with $\eta^{(+)}$ ($\eta^{(-)}$)
corresponding to the 
part containing annihilation (creation) operators in the expansion in terms of
plane waves.
Consequently, using similar arguments as before, the expectation value of
$\pa_{\pm} \eta $ vanishes on such states, 
\be
\langle \Psi_0 \mid \pa_{\pm} \eta \mid \Psi_0 \rangle  = 0 
\lab{reduceetab}
\ee
In addition, it follows that
\be
\langle \Psi_0 \mid e^{ \eta} \mid \Psi_0 \rangle  = {\rm const.} 
\lab{reduceetac}
\ee

Notice that the operator $e^{ \eta}$ has conformal weights $\( -\h , -\h\)$,
and since it is present in the Lagrangian \rf{lagraboson2}, it follows that 
such reduction does break the conformal symmetry. Such breaking of symmetry
resembles the Higgs mechanism since it generates  mass for the $\phi$ field,
and that is  
proportional to the expectation value of $e^{ \eta}$ \ct{hir}. 

After the reductions \rf{reduce} and \rf{reduceeta}, the $\nu$ field decouples
and we are left with a unitary theory which is that of the sine-Gordon model
of the $\phi$ field. In fact, as Coleman has shown \ct{coleman}, the
sine-Gordon theory is unbounded below if $\b^2 > 8 \pi$. Therefore, from
\rf{betadef} we have to have
\be
\b^2 < 8 \pi \qquad \ra \qquad k > {\mid \pi -2 e_{\psi}\mid \o 2 \pi} 
\ee

An important property of the model under consideration is that it
presents another spinor field which is not confined. Indeed, using bosonization
rules again, we can introduce a spinor $\chi$ as 
\br
i : {\bar \chi} \gamma^{\mu} \pa_{\mu} \chi : &=& \frac{\b^2}{8\pi}\, 
:\( \pa \phi\)^2 :\\
: {\bar \chi} \chi : &=& -\frac{{\bar \mu}}{2\pi} \, : \cos \( \b \phi \) :\\
:{\bar \chi} \gamma^{\mu} \chi : &=& - {\b\o 2\pi}\, \epsilon^{\mu\nu}
\pa_{\nu} \, \phi
\er

The Lagrangian \rf{lagraboson2} becomes (assuming \rf{condepsi}) 
\br
 \cl = - {1\o 2} \( \pa_{\mu} \rho \)^2 
+ \h  \pa_{\mu} \nu \, \pa^{\mu} \eta
- {k\o 8}\, m_{\psi}^2 \, e^{2\,\eta} 
+ e_{\psi}\( i  {\bar \chi} \gamma^{\mu} \pa_{\mu} \chi 
- m_{\chi}\, e^{\eta} \, {\bar \chi} \chi 
- \frac{g}{2} \, \( {\bar \chi} \gamma^{\mu} \chi \)^2 \) 
\lab{lagraboson3}
\er 
with
\be
\frac{4\pi}{\b^2} = 1 + \frac{g}{\pi} \qquad \qquad 
m_{\chi} = \frac{\mu}{{\bar \mu}} \, k \, m_{\psi}
\ee

Therefore, we get a Thirring model coupled to $\eta$ plus the free scalar
field  $\rho$. After the reductions \rf{reduce} and \rf{reduceeta} the theory 
becomes the pure massive Thirring model. Notice that $\chi$ becomes a free
massive spinor \ct{coleman} for $\b^2 = 4 \pi$ or 
$k = \mid \pi -2e_{\psi}\mid /\pi$. 

The properties of the theory \rf{lagrangian} at the quantum level are quite
remarkable. In the weak couling regime, i.e. small $k$, the excitations around
the vacuum correspond to the spinor $\psi$ and the ``gauge''particle
$\vp$. The $U(1)$ symmetry \rf{globalu1} is not broken and the charged states
correspond to the $\psi$ particles. Consider now those states satisfying
\rf{reduce} and look for the fluctuations around the state corresponding, for
instance, to the one-soliton solution \rf{solsimple}. The $\psi$
particles disappear from the spectrum since they are confined inside the
soliton. The $\psi$ particles can live outside the soliton only in bound
states with vanishing $U(1)$ charge. The theory, however, presents another
spinor particle corresponding to the excitations of the Thirring field $\chi$,
which have zero $U(1)$ charge. However, according to Coleman's interpretation
of the sine-Gordon/Thirring equivalence, such excitations correspond to the
solitons themselves. Therefore, we can make an analogy with what is expected
to happen in QCD. The $\psi$ and $\chi$ particles are like the quarks and 
 hadrons respectively. The $U(1)$ charge is analogous to color in QCD, since
it is also confined. 

Another important point, first observed in \ct{witten}, is that although the
theory \rf{lagrangian} presents the chiral symmetry \rf{chiraltransf}, it
does present massive spinor states. The reason is quite simple and elegant. 
Using the bosonization rule \rf{boson3} and the field redefinitions 
\rf{redeffields}, one gets that the chiral current \rf{chiral} can be written
as 
\be
J^{\mu}_5 =  \sqrt{\frac{\mid  \pi - 2 e_{\psi} \mid}{2 \pi}}\,\, 
\pa^{\mu} \rho
\ee
Therefore, the spinor $\chi$ does not contribute to such current and so has
zero chirality. Therefore, the fact that it does acquire a mass is not
incompatible with the chiral symmetry. The spinor with non zero chirality is
$\psi$ and it disappears from the spectrum in the confining sector of the
theory. The possibility of having chiral symmetry and massive fermions is
another  remarkable property of the theory \rf{lagrangian}. 

For these reasons the model \rf{lagrangian} constitute an excellent laboratory
to test ideas about confinement, the role of solitons in quantum field
theories, and duality transformations interchanging the role of solitons and
fundamental particles. With that motivation we shall now study the two-soliton
solutions and their interactions.

\section{Soliton solutions}
\label{sec:soliton}

It is well known the relevance of localized classical solutions of
non-linear relativistic field equations to the corresponding quantum
theories. In particular, solitons can be associated with quantum
extended-particle states. 
Therefore, we will examine the classical soliton type solutions to get
insight into the quantum spectrum of the model, in much the same way of what is
already known in the remarkable sine-Gordon model. We argue that, at the
classical level, the solutions for the $\varphi $ and $\psi $ fields share some
features of the sine-Gordon and the massive Thirring theories, respectively. 
We have shown in the previous section, using bosonization methods,  that at
the quantum level the 
theory \rf{lagrangian} 
is equivalent to \rf{lagraboson2}.  However, the space of classical solutions
of the two theories are not the same. One can not use 
\rf{boson1}-\rf{boson3}, at the classical level, as change of variables to
relate classical solutions of one theory to those of the other. The
bosonization rules  \rf{boson1}-\rf{boson3} are purely quantum relations. 

The one-soliton solutions of \rf{lagrangian} have been calculated in
\ct{matter} using the dressing transformation method. Here, we construct the
two-soliton solutions using the same methods. We start with the zero curvature
representation of the equations of motion of the theory  
\be 
\sbr{\pa_{+} + A_{+}}{\pa_{-} + A_{-}}=0
\lab{zc}
\ee
where the indices $\pm$ stand for the light cone variables $x_{\pm} = t \pm
x$. For the  theory \rf{lagrangian} the connections $A_{\pm}$ live on the
$\widehat{sl} (2)$ affine Kac-Moody algebra $\cgh$, and are given in the
appendix 
\ref{appa} (see \rf{gp}). Since \rf{zc} imply the connections must be flat,
one can write them  
as $A_{\mu} = - \pa_{\mu} T \, T^{-1}$, with $T$ being a group element
obtained by exponentiating $\cgh$. In addition, using an integral  gradation
of $\cgh$, i.e. $\cgh = \oplus_{n} \cgh_n$, $n \in \IZ$, one can perform a
generalized Gauss decomposition of the element $T \rho T^{-1}$, with $\rho$
being  a given constant group element. 
Denote by $\cgh_{<0}$, $\cgh_0$ and $\cgh_{ >0}$ the subalgebras generated by
elements of 
grades negative, zero and positive respectively. Then 
\be
T \rho T^{-1} = \( T \rho T^{-1} \)_{<0} \, \( T \rho T^{-1} \)_{0} \, 
\( T \rho T^{-1} \)_{>0}
\ee
where $\( T \rho T^{-1} \)_{<0} $, $\( T \rho T^{-1} \)_{0}$  and $\( T \rho
T^{-1} \)_{>0}$ are elements belonging to subgroups whose algebras are 
$\cgh_{<0}$, $\cgh_0$ and $\cgh_{ >0}$ respectively. One now introduces 
\be
T^{\rho} \equiv \( T \rho T^{-1} \)_{>0} \, T = 
\( T \rho T^{-1} \)_{0}^{-1} \, \( T \rho T^{-1} \)_{<0}^{-1} \, T \rho 
\lab{trho}
\ee
Such relation defines a transformation on the connections 
\be
A_{\mu} = - \pa_{\mu} T \, T^{-1} \qquad \ra \qquad 
A_{\mu}^{\rho} = - \pa_{\mu} T^{\rho} \, {T^{\rho}}^{-1}
\lab{pottransf}
\ee
which preserves their grading structure. Therefore, if one knows a solution
$T$ for the zero curvature one obtains a new solution $T^{\rho}$, determined
by the constant group element $\rho$.  We point out that the fact that the
transformed element $T^{\rho}$ can be written in two different ways as in
\rf{trho} plays a crucial role in the dressing method. It guarantees that the
transformed connection is in the same gauge as the original one, and therefore
allows to translate the dressing transformation \rf{pottransf} into
transformation of the physical fields defining the theory.   

A quite general procedure to construct soliton solutions in integrable
theories, using such dressing transformations,  is described in \ct{tau}. 
It constitutes a generalization of the so-called ``solitonic specialization''
in the context of the Leznov-Saveliev solution for Toda type models
\ct{otu,ous,ls}.  
One starts by looking for a ``vacuum'' solution 
such that the connections $A_{\pm}$, when evaluated on it, belong to an algebra
of oscillators, i.e. an abelian
(up to central terms) subalgebra of the Kac-Moody algebra. One then looks for
the eigenvectors $V_i$, in $\cgh$, of such oscillators. The solitons belong to
the orbits of solutions obtained by the dressing transformation performed by
elements of the form $\rho = e^{V_{i_1}} \, e^{V_{i_2}} \ldots e^{V_{i_n}}$.  

For the theory \rf{lagrangian} we perform the dressing transformation starting
from the vacuum solution
\be
\varphi =\psi =\widetilde{\psi }=\eta =0, \, \,  \quad 
\qquad 
{\nu }=-\frac{1}{8}m_{\psi}^{2}x_{+}x_{-}\equiv \nu_0 
\lab{nicevac}
\ee
The connection evaluated on such solution is given by (see \rf{gp})
\be
A_{+}^{\rm vac} = -  \, E_2 \qquad \qquad 
A_{-}^{\rm vac} = E_{-2} + \frac{1}{8} \, \, m_{\psi}^2 x_{+} C 
\ee
In addition, one has
\be
A_{\pm}^{\rm vac} = - \pa_{\pm } T_{\rm vac} \, T_{\rm vac}^{-1} 
\qquad {\rm with} \qquad 
T_{\rm vac} = e^{ x_{+} E_2} \, e^{-x_{-} E_{-2}}
\ee

Since $E_{\pm 2}= {1\o 4}m_{\psi} H^{\pm 1}$ (see  \rf{e2}) the relevant 
oscillators are $H^{\pm n}$ and their algebra is given by \rf{sl2a}. The
eigenvectors of the oscillators are given by 
\be
V_{\pm }(z)=\sum_{n\in \IZ}z^{-n}E_{\pm }^{n}
\ee
and
\br
\lb E_{2}\, ,\, V_{\pm}(z) \rb &=& \pm \h m_{\psi} z \, V_{\pm}(z) , \nonu \\
\lb E_{-2}\, ,\, V_{\pm}(z) \rb &=& \pm \h { m_{\psi}\o z }\, V_{\pm}(z)
\lab{eigeneq}
\er

Notice that $V_{+}\( z\)$ and $V_{-}\( -z\)$ have the same eigenvalues, and 
such degeneracy is related to the global $U(1)$ symmetry discussed in
\rf{globalu1} \ct{matter}.   

The solutions in the orbit of the vacuum \rf{nicevac}, under the dressing
transformations, are given by \ct{matter}
\be
e^{-i\varphi }=\frac{\tau _{1}}{\tau _{0}} 
,\qquad \qquad \qquad e^{-(\nu -\nu_{o} -{i\o 2} \vp)}=\tau _{0}
\lab{tauphinu}
\ee
and
\br
\psi = \sqrt{\frac{m_{\psi}}{4i}} \, \( 
\begin{array}{r}
\tau_{R}/\tau_0\\
-\tau_{L}/\tau_1
\end{array} \) \, \, , \qquad \qquad 
\widetilde{\psi }= -\sqrt{\frac{m_{\psi}}{4i}} \, \( 
\begin{array}{r}
\widetilde{\tau}_{R}/\tau_1\\
\widetilde{\tau}_{L}/\tau_0
\end{array} \) 
\lab{taupsi}
\er
where we have introduced the tau-functions
\br
\tau_0 &\equiv& <\widehat{\lambda }_{0}\mid G\mid \widehat{\lambda }_{0}>
\, , \qquad \; \; \; \; \; \; \; 
\tau_1 \equiv <\widehat{\lambda }_{1}\mid G\mid \widehat{\lambda }_{1}>
\nonu\\ 
\tau_{R}&\equiv& <\widehat{\lambda }_{0}\mid E_{-}^{1}G\mid 
\widehat{\lambda }_{0}> 
\, , \qquad \; 
\widetilde{\tau}_{R}\equiv <\widehat{\lambda }_{1}\mid
E_{+}^{0}G\mid\widehat{\lambda }_{1}> 
\nonu\\ 
\tau_{L}&\equiv& <\widehat{\lambda }_{1}\mid GE_{-}^{0}\mid \
\widehat{\lambda }_{1}>
\, , \qquad 
\widetilde{\tau}_{L}\equiv <\widehat{\lambda }_{0}\mid GE_{+}^{-1}\mid
\widehat{\lambda }_{0}>
\lab{taus}
\er
and where
\be
G \equiv T_{\rm vac} \, \rho \, T_{\rm vac}^{-1} = 
 e^{ x_{+} E_2} \, e^{-x_{-} E_{-2}}  \, \rho \,  
e^{x_{-} E_{-2}}  \,  e^{ -x_{+} E_2} 
\lab{G}
\ee
We have denoted $\mid  \widehat{\lambda }_{0}>$ \ and 
$\mid  \widehat{\lambda}_{1}>$  the highest weight 
states of the two fundamental representations of the affine Kac-Moody algebra 
$\widehat{sl}(2),$ respectively the scalar and spinor ones. They satisfy
\be
H^0 \, \mid  \widehat{\lambda }_{0}> = 0 \, , \qquad 
H^0 \, \mid  \widehat{\lambda}_{1}> = \mid  \widehat{\lambda}_{1}>\, , \qquad 
C \mid  \widehat{\lambda}_{j}> = \mid  \widehat{\lambda}_{j}> 
\ee
with $j=0,1$, and in addition 
\be
E_{+}^0 \, \mid \widehat{\lambda}_{j}> = H^n \, \mid  \widehat{\lambda}_{j}> = 
E^n_{\pm} \,  \mid \widehat{\lambda}_{j}> =  0 \, , 
\qquad \qquad \mbox{\rm for $n>0$}
\ee

Notice that if one makes the shift $\rho \ra h \rho h^{-1}$, with $h =
e^{i\theta  H^0/2}$, one gets that $G \ra h G h^{-1}$, since $h$ commutes
with $E_{\pm 2}$. Therefore, the tau-functions transform as
\be
\tau_0 \ra \tau_0 \, , \quad \tau_1 \ra \tau_1 \, , \quad 
\tau_{R} \ra e^{i\theta} \, \tau_{R} \, , \quad 
\tau_{L} \ra e^{i\theta} \, \tau_{L} \, , \quad 
\widetilde{\tau}_{L} \ra e^{-i\theta} \, \widetilde{\tau}_{L} \, , \quad 
\widetilde{\tau}_{R} \ra e^{-i\theta} \, \widetilde{\tau}_{R} 
\lab{globalu1tau}
\ee
which corresponds to the global $U(1)$ transformations \rf{globalu1}. 

We will be interested in classical solutions satisfying the relations 
\rf{equivcurrents}. In terms of the tau-functions defined above those
relations can be written as
\br
\tau_0 \pa_{+} \tau_1 -\tau_1 \pa_{+} \tau_0 &=& 
\h \, m_{\psi}\, \tau_{R} \, \widetilde{\tau}_{R} \nonu\\ 
\tau_0 \pa_{-} \tau_1 -\tau_1 \pa_{-} \tau_0 &=& 
\h\, m_{\psi}\, \tau_{L} \, \widetilde{\tau}_{L}
\lab{equivcurtau}
\er

Using \rf{taus}-\rf{G} one can write \rf{equivcurtau} as the algebraic
relations 
\br
<  G >_{{0}} \,
< H^1\, G>_{{1}} - 
<  G >_{{1}}\, 
< H^1\, G>_{{0}}
&=&2 < E_{-}^{1}G>_{{0}}  \,
< E_{+}^{0}G>_{{1}} 
\nonu \\
& & \lab{equivalgrel}\\
<  G>_{{0}} \, 
<  G \, H^{-1}>_{{1}} - 
< G>_{{1}} \, 
<  G\, H^{-1}>_{{0}}
&=& 2 < GE_{+}^{-1}>_{{0}} \, 
< GE_{-}^{0}>_{{1}}
\nonu 
\er
where we have denoted 
$<\widehat{\lambda }_{a} \mid X \mid \widehat{\lambda }_{a} > 
\equiv < X >_{a}$. 
These equations  determine the group elements $\rho$ that lead to solutions
satisfying the relations \rf{equivcurrents}. 

One can use \rf{tauphinu}-\rf{taupsi} to write the equations of motion of the
theory \rf{lagrangian} in terms of the tau-functions. The equations of motion 
\rf{sl2eqm1}-\rf{sl2eqm2} for $\vp$ and $\nu$ imply that (for $\eta =0$)
\br
\tau_0 \pa_{+} \pa_{-} \tau_0 - \pa_{+} \tau_0 \, \pa_{-} \tau_0 &=& 
\frac{1}{4} \, m_{\psi}^2 \, \tau_{L}\, \widetilde{\tau}_{R} 
\lab{tau2order1}\\
\tau_1 \pa_{+} \pa_{-} \tau_1 - \pa_{+} \tau_1 \, \pa_{-} \tau_1 &=& 
\frac{1}{4} \, m_{\psi}^2 \, \tau_{R}\, \widetilde{\tau}_{L} 
\lab{tau2order2}
\er
The equations of motion \rf{sl2eqm4}-\rf{sl2eqm5} for the Dirac spinors imply
(for $\eta =0$) 
\br
\tau_0 \pa_{-} \tau_{R} - \tau_{R} \pa_{-} \tau_{0} &=& 
- \h \, m_{\psi} \, \tau_{1}\, \tau_{L} 
\lab{tau1order1}\\
\tau_1 \pa_{+} \tau_{L} - \tau_{L} \pa_{+} \tau_{1} &=& 
 \h \, m_{\psi} \, \tau_{0}\, \tau_{R} 
\lab{tau1order2}\\ 
\tau_1 \pa_{-} \widetilde{\tau}_{R} - \widetilde{\tau}_{R} \pa_{-} \tau_{1} 
&=& 
 \h \, m_{\psi} \, \tau_{0}\, \widetilde{\tau}_{L} 
\lab{tau1order3}\\
\tau_0 \pa_{+} \widetilde{\tau}_{L} - \widetilde{\tau}_{L} \pa_{+} \tau_{0} 
&=& 
- \h \, m_{\psi} \, \tau_{1}\, \widetilde{\tau}_{R} 
\lab{tau1order4}
\er
The relations \rf{tau2order1}-\rf{tau1order4} are the Hirota's bilinear
equations \ct{hirota} for the model \rf{lagrangian}. 
Notice that the first order equations \rf{equivcurtau} and
\rf{tau1order1}-\rf{tau1order4} imply that the difference of \rf{tau2order1}
and \rf{tau2order2} should be satisfied, but not that each one separately
should do.  

The $N$-soliton solutions are constructed by taking the constant group element
$\rho$ of the dressing transformation to be a product of exponentials of the
eigenvectors of $E_{\pm 2}$ (see \rf{eigeneq}). Since $V_{+}(z)$ and
$V_{-}(-z)$ have the same eigenvalue, we take the exponential of a linear
combination of them, i.e. 
\be
\rho_{\rm N-sol} \equiv \prod_{l=1}^N \, e^{V(a_{\pm}^{(l)},z_l)}
\qquad {\rm with} \qquad 
V(a_{\pm}^{(l)},z_l) \equiv 
\sqrt{i}\left ( a_{+}^{(l)} V_{+}(z_l) + a_{-}^{(l)} V_{-}(-z_l)\right);
\lab{nsolitsol}
\ee
In fact, if for a given $l$ (or more than one) either $a_{+}^{(l)}$ or
$a_{-}^{(l)}$ vanishes 
one does not obtain $N$-soliton solutions. For example, in the case of the
one-soliton one gets  a vanishing  solution for 
$\vp$ \ct{matter}. The fact that the
$N$-soliton solutions need the two degenerate eigenvectors, i.e. 
$V_{+}(z)$ and $V_{-}(-z)$,  is what makes them to
carry, besides the topological charge, the ``electric'' $U(1)$ charge
\rf{noethersl2}.

The corresponding group element \rf{G} becomes
\be
G_{\rm N-sol} = \prod_{l=1}^N 
\,\exp\( e^{\Gamma (z_l)} \,V(a_{\pm}^{(l)},z_l) \) 
\lab{GNsol}
\ee
with 
\be
\Gamma (z_l)=\h \, m_{\psi} \, (z_l x_{+}-\frac{1}{z_l}x_{-}) \equiv 
\gamma_l \( x - v_l t\)  
\lab{gammadef}
\ee 
and so
\be
\gamma_l = \h \, m_{\psi} \, \( z_l + \frac{1}{z_l}\) 
= ({\rm sign} \, z_l) \; \frac{ m_{\psi}}{\sqrt{1 - v^2_l}}
\qquad \qquad 
v_l = \frac{1- z^2_l}{1+ z^2_l} 
\lab{gammaandv}
\ee

Notice that one needs to take $z_l$ real for the soliton velocities to be
smaller than light speed, i.e. $\mid v_l \mid \leq  1$. 
In fact, the quantities $z_l$ are related to the
rapidities $\theta_l$ of the solitons through 
\be
z_l \equiv \epsilon_l \, e^{\theta_l}
\lab{zthetarel}
\ee
with $\epsilon_l = \pm 1$, and so from \rf{gammadef} 
\be
\Gamma (\theta_l )= \epsilon_l \, m_{\psi} 
\( x \cosh \theta_l + t \sinh \theta_l \) 
\ee

An important point in the  calculations involved in the construction of the
soliton solutions is that it is much easier to work with
the homogeneous (or Fubini-Veneziano) vertex operator construction for
$V_{\pm}(z)$. Indeed, in  
such construction one has that\footnote{See eq. (14.8.14) on page 309 of
ref. \ct{kac}, or eq. (6.2.6) of ref. \ct{go} for the details.}
\be
V_{+} \( z \)  V_{+} \( \zeta \) \ra 0 \, , \qquad 
V_{-} \( z \)  V_{-} \( \zeta \) \ra 0 \, , \qquad \quad 
\mbox{\rm as $z \ra \zeta$}
\ee
and therefore the exponentials involving  $V_{\pm} \( z \)$ truncate.

We are interested in the solutions where the field $\vp$ is real. 
Using \rf{tauphinu}  one obtains that (since $-i\vp = \log \mid
\frac{\tau_1}{\tau_0}\mid  + i\, {\rm arg}\, \frac{\tau_1}{\tau_0}$) 
\be
\vp^* = \vp \qquad {\rm implies } \qquad \mid \tau_0 \mid = \mid \tau_1 \mid 
\lab{realcond1}
\ee
and consequently 
\be
\vp = \arctan \( i\,
\frac{\tau_0^* \tau_1 - \tau_0 \tau_1^*}{\tau_0^* \tau_1 + \tau_0 \tau_1^*}\) 
\; , \qquad \qquad {\rm or} \qquad \qquad 
\vp = \zeta_0 - \zeta_1 + n \pi 
\lab{realvp}
\ee
where $n$ is any integer, and where we have denoted $\tau_a= \mid \tau_a
\mid\, e^{i \zeta_a}$, $a=0,1$.  

Using  \rf{taupsi} one observes that the reality condition \rf{psitrealcond}
implies that the tau-functions have to satisfy 
\be
\frac{\widetilde{\tau}_{R}}{\tau_1} = 
- i e_{\psi}\,\frac{\tau_{R}^*}{\tau_0^*}  
\; , \qquad \qquad \quad 
\frac{\widetilde{\tau}_{L}}{\tau_0}= 
i e_{\psi}\,\frac{\tau_{L}^*}{\tau_1^*}
\lab{taurealcond}
\ee

\subsection{The one-soliton solutions}

The one-soliton solution can easily be calculated by evaluating the matrix
elements \rf{taus} with $\rho = e^{V(a_{\pm},z)}$, or alternatively by
applying the Hirota's method to the equations
\rf{tau2order1}-\rf{tau1order4}. The result is 
\begin{eqnarray}
\tau_0 &=& 1 - {i\over 4}\,a_{-}\,a_{+}\,e^{2\Gamma ( z)}\, \; ; 
\qquad 
\tau_1 = 1 + {i\over 4}\,a_{-}\,a_{+}\,e^{2\Gamma ( z)}\,
\nonumber\\
\tau_R &=&  \sqrt{i}\,a_{+}\,\,z\,e^{\Gamma (z)} \; ; 
\qquad \qquad \; 
{\widetilde \tau}_R = \sqrt{i} \,a_{-}\,e^{\Gamma (z)}
\nonumber\\
\tau_L &=&  \sqrt{i}\,a_{+}\,e^{\Gamma (z)}\, \; ; 
\qquad \qquad \quad 
{\widetilde \tau}_L = - \sqrt{i} \,\frac{a_{-}}{z}\, e^{\Gamma (z)}
\lab{tauonesol}
\end{eqnarray}
One can check that such solutions also satisfy \rf{equivcurtau} without any
further restriction. Therefore, the one-soliton solution satisfy the condition
\rf{equivcurrents} for the equivalence between Noether and topological
currents. 

If one requires that $\vp$ should be real, then one gets from \rf{realcond1}
and \rf{tauonesol} that $\( a_{-}\,a_{+}\)$ must be real. Such condition is
sufficient to make $\tau_1 = \tau_0^*$. Therefore, using 
\rf{topological} and \rf{realvp}, one gets that the topological charge of
the soliton is 
\br 
Q_{\rm topol.} &=& -\, {\rm sign}\, \( a_{-}\,a_{+}\)  \qquad  \qquad 
\mbox{\rm for $z > 0$}\nonu\\
Q_{\rm topol.} &=&  {\rm sign}\, \( a_{-}\,a_{+}\)  \qquad  \qquad \quad 
\mbox{\rm for $z < 0$} 
\lab{onesoltopovpreal}
\er

However, if one imposes in addition, the reality condition \rf{psitrealcond}
on the spinors 
one gets from \rf{tauonesol} that the parameters must satisfy
\be
a_{-} = - e_{\psi} \, a_{+}^* \, z 
\lab{pararealcond}
\ee
The topological charge in such case is
\be
Q_{\rm topol.} =  {\rm sign}\, e_{\psi} \qquad  \qquad \qquad 
\mbox{\rm for any real $z$} 
\lab{onesoltopovppsireal}
\ee

Notice that in
general, solitons are transformed into anti-solitons by parity
transformations. However, the theory \rf{lagrangian} is not invariant under
spatial parity, and the CP-like symmetry \rf{cpsymmetry} is broken when the
reality condition \rf{psitrealcond} is imposed. In order to have a degenerate
vacua and topological solitons we need the factor $i$ in the exponential of
the potential term in \rf{lagrangian}. That makes the Lagrangian complex and
so it is very unlikely that \rf{lagrangian} has a CPT symmetry. The lack of P,
CP and CPT symmetries is what reaffirms  the existence of only the soliton or
anti-soliton solution for a given choice of $e_{\psi}$ in  \rf{psitrealcond}.

We then reach an interesting conclusion: without the reality condition
\rf{psitrealcond} the theory \rf{lagrangian} has two Dirac spinors and it
also has the soliton and anti-soliton solutions, since according to
\rf{onesoltopovpreal} both signs of the charge are admissible. However, if we
impose the  reality condition 
\rf{psitrealcond} the theory \rf{lagrangian} looses one Dirac spinor and
also one soliton solution, since for a given choice of $e_{\psi}$ one observes
from \rf{onesoltopovppsireal} that only one topological charge (corresponding
to soliton or anti-soliton) is permitted.  That is indicative of the existence
of some sort of duality between solitons and spinor particles.

The particular one-soliton solution \rf{solsimple} satisfies
\rf{pararealcond} with $e_{\psi} =1$, $z>0$, $\theta$ being the phase of
$a_{+}$, i.e. $a_{+} = \mid a_{+}\mid e^{i\theta}$, and 
$\mid a_{+}\mid \sqrt{z}/2 = \exp \( - m_{\psi}x_0 /\sqrt{1-v^2}\)$.   

The mass of these one-soliton solutions was evaluated in \ct{matter} and it is 
given by
\be
M = 2\, k\, m_{\psi}
\lab{solmass}
\ee
where $k$ is the coupling constant appearing in the Lagrangian
\rf{lagrangian}. 

\subsection{The two-soliton solutions}

The two-soliton solutions are calculated through the dressing transformation
method taking  (see \rf{nsolitsol}) 
\be
\rho_{\rm 2-sol} \equiv e^{V(a^{(1)}_{\pm},z_1 )} \, e^{V(a^{(2)}_{\pm},z_2)} \qquad 
\mbox{\rm and so} \qquad 
G_{\rm 2-sol} = e^{(e^{\Gamma (z_1)}V(a^{(1)}_{\pm},z_1))} \, 
e^{(e^{\Gamma (z_2)}V(a^{(2)}_{\pm},z_2))}
\ee

The explicit solution is calculated by evaluating the matrix elements
\rf{taus}. Alternatively, one can apply Hirota's expansion method \ct{hirota}
to the Hirota's equation \rf{tau2order1}-\rf{tau1order4}. The results one
obtains are 
\begin{eqnarray}
\tau_0&=& 1 - {i\over 4}\,a^{(1)}_{-}\,a^{(1)}_{+}\,
   \,{{e^{2\Gamma (z_1)}}} - {i\over 4}\,a^{(2)}_{-}\,a^{(2)}_{+}\,
   \,{e^{2\Gamma (z_2)}} 
\nonumber\\ 
&-& i\,\frac{z_1\,z_2}{\left( z_1 + z_2 \right)^2}
\left( a^{(1)}_{+}\,a^{(2)}_{-} + a^{(1)}_{-}\,a^{(2)}_{+} \right) \,
      \,e^{\Gamma (z_1)+\Gamma (z_2)}
\nonumber\\    
&-& \frac{1}{16}\, 
\frac{\left( z_1 - z_2 \right)^4}{\left( z_1 + z_2 \right)^4}\, 
a^{(1)}_{-}\,a^{(1)}_{+}\,a^{(2)}_{-}\,a^{(2)}_{+}\,\,e^{2\( \Gamma (z_1)+
   \Gamma (z_2)\)}   
\lab{soltau0}    \\
\tau_1&=& 1 + {i\over 4}\,a^{(1)}_{-}\,a^{(1)}_{+}\,
   \,{{e^{2\Gamma (z_1)}}} + {i\over 4}\,a^{(2)}_{-}\,a^{(2)}_{+}\,
   \,{e^{2\Gamma (z_2)}} 
\nonumber\\ 
&+& \frac{i}{\left( z_1 + z_2 \right)^2}
\,\, \left( a^{(1)}_{+}\,a^{(2)}_{-}\,{{z_1}^2} +
   a^{(1)}_{-}\,a^{(2)}_{+}\,{{z_2}^2} \right) 
e^{\Gamma (z_1)+\Gamma (z_2)}
\nonumber\\
&-& \frac{1}{16}\, 
\frac{\left( z_1 - z_2 \right)^4}{\left( z_1 + z_2 \right)^4}
a^{(1)}_{-}\,a^{(1)}_{+}\,a^{(2)}_{-}\,a^{(2)}_{+}\,\,e^{2\( \Gamma
   (z_1)+\Gamma (z_2)\)} 
\lab{soltau1} 
\\ 
\frac{\tau_{R}}{\sqrt{i}}&=& \,a^{(1)}_{+}\,\,z_1\,e^{\Gamma (z_1)} 
+ \,a^{(2)}_{+}\,\,z_2 \,e^{\Gamma (z_2)}
- \frac{i}{4}\,
\frac{z_2 \,\left( z_1 - z_2 \right)^2}{\left( z_1 + z_2 \right)^2}\, 
\,a^{(1)}_{-}\,a^{(1)}_{+}\,a^{(2)}_{+}\,\,e^{2\Gamma (z_1)+\Gamma (z_2)}
\nonumber\\
&-& 
  \frac{i}{4}\, 
\frac{z_1\,\left( z_1 - z_2 \right)^2}{\left( z_1 + z_2 \right)^2}
\,a^{(1)}_{+}\,a^{(2)}_{-}\,a^{(2)}_{+}\,e^{\Gamma (z_1)+2\Gamma (z_2)}
\lab{soltaur} \\
\frac{\tau_{L}}{\sqrt{i}}&=& a^{(1)}_{+}\,e^{\Gamma (z_1)} 
+ a^{(2)}_{+}\,e^{\Gamma (z_2)} 
+ \frac{i}{4} \, 
\frac{\left( z_1 - z_2 \right)^2}{\left( z_1 + z_2 \right)^2}\, 
a^{(1)}_{+}\,a^{(2)}_{-}\,a^{(2)}_{+}\,\,
      e^{\Gamma (z_1)+2\Gamma (z_2)}
\nonumber\\
&+& \frac{i}{4} \, 
\frac{\left( z_1 - z_2 \right)^2}{\left( z_1 + z_2 \right)^2}\, 
a^{(1)}_{-}\,a^{(1)}_{+}\,a^{(2)}_{+}\,e^{2\Gamma (z_1)+\Gamma (z_2)}
\lab{soltaul} \\
\frac{\widetilde{\tau}_{R}}{\sqrt{i}}&=& a^{(1)}_{-}\, e^{\Gamma (z_1)} 
+ a^{(2)}_{-}\,e^{\Gamma (z_2)} 
+ \frac{i}{4} \, 
\frac{\left( z_1 - z_2 \right)^2}{\left( z_1 + z_2 \right)^2}
\,a^{(1)}_{-}\,a^{(1)}_{+}\,a^{(2)}_{-}\,\, e^{2\Gamma (z_1)+\Gamma (z_2)}
\nonumber\\
&+& \frac{i}{4} \, 
\frac{\left( z_1 - z_2 \right)^2}{\left( z_1 + z_2 \right)^2}
  \,a^{(1)}_{-}\,a^{(2)}_{-}\,a^{(2)}_{+}\,e^{\Gamma (z_1)+2\Gamma (z_2)}
\lab{soltaurt} \\
\frac{\widetilde{\tau}_{L}}{\sqrt{i}}&=& -\frac{a^{(1)}_{-}}{z_1}\,\,e^{\Gamma (z_1)} 
- \frac{a^{(2)}_{-}}{z_2}\,e^{\Gamma (z_2)}
+ \frac{i}{4} 
\frac{\left( z_1 - z_2 \right)^2}{z_1\,\left( z_1 + z_2 \right)^2}
a^{(1)}_{-}\,a^{(2)}_{-}\,a^{(2)}_{+}\,
      e^{\Gamma (z_1)+2\Gamma (z_2)}
\nonumber\\
&+& \frac{i}{4}\, 
\frac{\left( z_1 - z_2 \right)^2}{z_2\,\left( z_1 + z_2 \right)^2}\, 
a^{(1)}_{-}\,a^{(1)}_{+}\,a^{(2)}_{-}\, e^{2\Gamma (z_1)+\Gamma (z_2)}
\lab{soltault} 
\end{eqnarray}

These tau-functions satisfy the Hirota's equations
\rf{tau2order1}-\rf{tau1order4} for any value of the constants $a^{(1)}_{\pm}$,
$a^{(2)}_{\pm}$, $z_1$ and $z_2$, which can be even complex. It is interesting
to 
notice that these tau-functions also satisfy the equivalence between the
Noether and topological currents \rf{equivcurtau}(or equivalently
\rf{equivcurrents}) without any further restrictions. 

As we have said the Lorentz transformations in $2d$ are $x_{\pm} \ra 
{\hat x}_{\pm} \equiv \l^{\pm 1} \, x_{\pm}$. The corresponding field
transformations can be 
obtained from \rf{ct}-\rf{ctf} by taking $f\(x_{+} \) = \l x_{+}$ and  
$g\(x_{-} \) = \l^{-1} x_{-}$. Therefore, choosing the conformal weight $\d$
of the field $\nu$ to vanish, one conclude from  \rf{tauphinu}-\rf{taupsi}
that, under Lorentz transformations, the tau-functions transform as
\be
\tau_a \( x\) \ra \tau_a \( {\hat x}\) =\tau_a \( x\) \qquad  \mbox{\rm for
$a=0,1$}
\ee
and
\be
\tau_R \( x\) \ra \tau_R \( {\hat x}\) = \l^{-\h}\, \tau_R \( x\) \qquad 
\tau_L \( x\) \ra \tau_L \( {\hat x}\) = \l^{\h}\, \tau_L \( x\)
\ee
and similarly for $\widetilde{\tau}_{R/L}$. One then observes that such
tranformations can be implemented in the space of solutions by transforming
the parameters of the solutions \rf{soltau0}-\rf{soltault} as
\be
z_i \ra \l \, z_i \qquad \qquad a^{(i)}_{\pm} \ra \l^{\mp \h} \,
a^{(i)}_{\pm} \qquad \qquad \qquad i =1,2 
\lab{lorentzpar}
\ee

In addition, the global $U(1)$ transformations \rf{globalu1} 
(see \rf{globalu1tau}) correspond in the space of parameters to 
\be
z_i \ra  z_i \qquad \qquad a^{(i)}_{\pm} \ra e^{\pm i \theta} \,
a^{(i)}_{\pm} \qquad \qquad \qquad i =1,2
\ee

\subsection{The reality conditions}

We want $z_1$ and $z_2$ real in order for the soliton velocities to be 
smaller than the speed of light ($c=1$, see \rf{gammaandv}). Then, one can
easily check that in order for the solutions \rf{soltau0}-\rf{soltau1} 
to satisfy \rf{realcond1} one has to have $a^{(1)}_{+} a^{(1)}_{-}$ and
$a^{(2)}_{+} a^{(2)}_{-}$ 
real, and so 
\be
a^{(i)}_{+} = \mid a^{(i)}_{+} \mid \, e^{i \xi_i}\, , \qquad 
a^{(i)}_{-} = \bareps_i \mid a^{(i)}_{-} \mid \, e^{-i \xi_i }\, , \qquad 
\qquad i=1,2 
\lab{realcond1a}
\ee
with $\xi_i$  real being phases, and $\bareps_i = \pm 1$ being the sign of
$a^{(i)}_{+} a^{(i)}_{-}$.  In addition, one needs
\be
\bareps_1\, z_1 \mid a^{(1)}_{+} \mid \, \mid a^{(2)}_{-} \mid = 
\bareps_2 z_2 \,\mid a^{(1)}_{-} \mid \, \mid a^{(2)}_{+}
\mid  
\lab{realcond1b}
\ee
Using \rf{zthetarel} one observes that \rf{realcond1b} implies
\br
\bareps_1\, \bareps_2\,  \eps_1 \, \eps_2 &=& 1 
\lab{realcond1b1}\\
e^{\theta_1 - \theta_2}\,  \mid a^{(1)}_{+} \mid \, \mid a^{(2)}_{-} \mid 
\, &=& \, 
\mid a^{(1)}_{-} \mid \, \mid a^{(2)}_{+}\mid  
\lab{realcond1b2}
\er

The relations \rf{realcond1a} and \rf{realcond1b} are the necessary and
sufficient conditions for \rf{realcond1} to be satisfied. However, they are
sufficient to make $\tau_1$ the complex conjugate of $\tau_0$. Therefore, 
\be
\vp^* = \vp \qquad \ra \qquad \tau_1 = \tau_0^* \qquad \ra \qquad 
\vp = 2 \zeta_0 + n \pi 
\lab{realvp2}
\ee
Consequently, the conditions \rf{taurealcond} become 
\be
\widetilde{\tau}_{R} = 
- i e_{\psi}\,\tau_{R}^* 
\; , \qquad \qquad \quad 
\widetilde{\tau}_{L}= 
i e_{\psi}\,\tau_{L}^*
\lab{taurealcond2sol}
\ee
One can easily check that the solutions \rf{soltau0}-\rf{soltault} satisfy the
conditions $\tau_1 = \tau_0^*$ (and so $\vp$ real) and \rf{taurealcond2sol} if
the parameters satisfy ($z_1$, $z_2$ being real) 
\be
a^{(i)}_{-} = - e_{\psi}\, {a^{(i)}_{+}}^* z_i \; \, 
\qquad \qquad  \quad i=1,2 
\lab{realcond2c}
\ee
Then it follows that the signs of $z_i$ and $a^{(i)}_{+} a^{(i)}_{-}$ are
related by
\be
\bareps_i = - \eps_i \, {\rm sign} \, e_{\psi} \qquad \qquad \qquad i=1,2 
\lab{realcond2c2}
\ee
which indeed satisfy \rf{realcond1b1}.

\subsection{The topological charges}

For the solutions satisfying the conditions \rf{realcond1a} and
\rf{realcond1b} we shall denote
\be
\mid a^{(i)}_{+} \mid \,  \mid a^{(i)}_{-} \mid \equiv 
e^{-2 \gamma_i x_0^{(i)}} \qquad \qquad \qquad i=1,2 
\ee
with $\gamma_i$ defined in \rf{gammaandv}. We then define 
\be
{\hat \Gamma} \( z_i \) \equiv \Gamma \( z_i \) - \gamma_i x_0^{(i)}
\ee
with $ \Gamma \( z_i \)$ being defined in \rf{gammadef}. 

In order to evaluate the topological charges we shall take the solution in the
center of mass reference frame, which corresponds to (see \rf{gammaandv} and
\rf{zthetarel}) 
\be
z_1 = \eps_1 e^{\theta}\; ; \qquad z_2 = \eps_2 e^{-\theta} 
\, , \qquad \ra \qquad 
v_1 = - v_2 \quad {\rm and} \quad \gamma_2 = \eps_1\, \eps_2 \, \gamma_1 
\lab{cmrf}
\ee
Then, using \rf{realcond1a}, \rf{realcond1b2},  \rf{cmrf}, and  
\rf{lorentzpar}, the
tau-function $\tau_0$ given in \rf{soltau0} becomes 
\br
\tau_0&=& 1 - \eps_1  \eps_2  \frac{\( \bareps_1 e^{\theta} - \bareps_2 
e^{-\theta}\)}{\( e^{\theta} + \eps_1 \eps_2  e^{-\theta} \)^2} 
\sin\(\xi_1 -\xi_2\)  e^{ {\hat \Gamma} (z_1)+ {\hat \Gamma} (z_2)}
- \frac{\bareps_1 \bareps_2}{16} 
\frac{\left( e^{\theta} - \eps_1 \eps_2  e^{-\theta}  
\right)^4}{\left(  e^{\theta} + \eps_1 \eps_2  e^{-\theta} \right)^4} 
e^{2\( {\hat \Gamma} (z_1)+ {\hat \Gamma} (z_2)\)} \nonu\\
&-& i \( \frac{\bareps_1}{4} \, e^{2 {\hat \Gamma} (z_1)} + 
\frac{\bareps_2}{4} \, e^{2 {\hat \Gamma} (z_2)} 
+ \eps_1 \, \eps_2 \, \frac{\( \bareps_1 e^{\theta} + \bareps_2 
e^{-\theta}\)}{\( e^{\theta} + \eps_1\, \eps_2  e^{-\theta} \)^2}\, 
\cos\(\xi_1 -\xi_2\) \, e^{ {\hat \Gamma} (z_1)+ {\hat \Gamma} (z_2)}\) 
\lab{tau0com}
\er   
In addition, one has $\tau_1 = \tau_0^*$, when \rf{realcond1b1} is also
enforced. Therefore, according to \rf{topological} and \rf{realvp2}, the
topological charge is $Q_{\rm topol.} = \frac{2}{\pi}\( {\rm arg}\, \tau_0 
\( \infty\) - {\rm arg}\, \tau_0 \( -\infty\)\)$. We shall use the
prescription that the ${\rm arg}\, \tau_0 $ lies between $-\pi$ and $\pi$. 
When evaluating the charge
one should notice that, on the center of mass reference frame,  
${\hat \Gamma} (z_1)+ {\hat \Gamma} (z_2)$ does not depend upon $x$ if
$\eps_1$ and $\eps_2$ have opposite signs.  In addition, for $\eps_1 \eps_2 =
\bareps_1 \bareps_2 =1$ one should pay attention on the sign of
the imaginary part of $\tau_0$ as it tends to zero, in order to determine if
${\rm arg}\, \tau_0 $ tends to $-\pi$ or $\pi$. In such case, one can write 
\be
{\rm Im} \tau_0 \ra  
- \h \bareps_1 \( 1  + \frac{\cos \( \xi_1 - \xi_2 \)}{\cosh \theta}  \) 
e^{2 m_{\psi} \eps_1 x \cosh\theta + \ldots } 
\qquad \qquad {\rm as} \quad x \ra \eps_1 \infty 
\ee
for $\eps_1 \eps_2 = \bareps_1 \bareps_2 =1$. Since $\frac{\cos \( \xi_1 -
\xi_2 \)}{\cosh \theta}\geq -1$, the limit is independent of $\xi_i$. In
fact, the topological charge, for $\vp$ real, does not depend upon the phases
$\xi_i$.  The results one obtains are
\br
Q_{\rm topol.} &=& 2 \bareps_1 \qquad \quad \mbox{\rm  for 
$\( \eps_1=\eps_2=-1, \bareps_1\bareps_2=1\) $ or  
$\( \eps_1=-\eps_2=-1, \bareps_1\bareps_2=-1\)$} \nonu \\
Q_{\rm topol.} &=& -2 \bareps_1 \quad \quad \mbox{\rm  for 
$\(\eps_1=\eps_2=1,\bareps_1\bareps_2=1\)$ or  
$\(\eps_1=-\eps_2=1,\bareps_1\bareps_2=-1\)$}
\lab{charge2solvpreal}
\er

If one now imposes the reality condition \rf{psitrealcond} on the spinors,
or equivalently \rf{taurealcond2sol}-\rf{realcond2c2}, one gets 
\be
Q_{\rm topol.} = 2 \, {\rm sign}\, e_{\psi} 
\lab{epsiconstr}
\ee
for any $\eps_i$ and $\bareps_i$ satisfying \rf{realcond2c2}. 

We then have a situation similiar to the one-soliton solutions, since in that
case  
the imposition of the condition \rf{psitrealcond} makes either the soliton or
anti-soliton to disappear from the spectrum. Here the same thing happens,
since \rf{psitrealcond} chooses  the sign of $e_{\psi}$ and then either the
charge $2$ or $-2$ solutions disapear. However, a new feature emerges. We saw
we could have (without \rf{psitrealcond}) soliton and anti-soliton solutions
with $\vp$ real. We have not found here a charge zero solution corresponding
to the scattering of a soliton and anti-soliton for $\vp$ real. Such solution
exists however, for $\vp$ complex but asymptotically real as we show below.

\subsubsection{Asymptotically real solutions}

One can easily verify that for the solutions \rf{soltau0}-\rf{soltault} the
field $\vp$ is always real asymptotically, i.e.  
$\mid \tau_0\mid \ra \mid \tau_1\mid$ as $x\ra \pm \infty$. In addition, 
one also has that $\psi, {\widetilde \psi} \ra 0$ as $x\ra \pm
\infty$. Therefore, the topological charge is always real for those
solutions. 

Let us now consider a solution of the type \rf{soltau0}-\rf{soltault} that
satisfies the conditions \rf{realcond1a} and \rf{realcond1b2} but not
\rf{realcond1b1} (and of course not \rf{psitrealcond} too). In the center of
mass reference frame (see \rf{cmrf}) $\tau_0$ is given by \rf{tau0com} and
$\tau_1$ by 
\br
\tau_1&=& 1 - \bareps_1  \bareps_2 
\frac{\( \bareps_1 e^{\theta} - \bareps_2 
e^{-\theta}\)}{\( e^{\theta} + \eps_1 \eps_2  e^{-\theta} \)^2} 
\sin\(\xi_1 -\xi_2\)  e^{ {\hat \Gamma} (z_1)+ {\hat \Gamma} (z_2)}
- \frac{\bareps_1 \bareps_2}{16}
\frac{\left( e^{\theta} - \eps_1 \eps_2  e^{-\theta}  
\right)^4}{\left(  e^{\theta} + \eps_1 \eps_2  e^{-\theta} \right)^4} 
e^{2\( {\hat \Gamma} (z_1)+ {\hat \Gamma} (z_2)\)} \nonu\\
&+& i \( \frac{\bareps_1}{4} \, e^{2 {\hat \Gamma} (z_1)} + 
\frac{\bareps_2}{4} \, e^{2 {\hat \Gamma} (z_2)} 
+ \bareps_1 \, \bareps_2 \, \frac{\( \bareps_1 e^{\theta} + \bareps_2 
e^{-\theta}\)}{\( e^{\theta} + \eps_1\, \eps_2  e^{-\theta} \)^2}\, 
\cos\(\xi_1 -\xi_2\) \, e^{ {\hat \Gamma} (z_1)+ {\hat \Gamma} (z_2)}\) 
\lab{tau1com}
\er   
Since $\mid \tau_0\mid = \mid\tau_1\mid$ holds true asymptotically, the
topological charge is then given by   
$Q_{\rm topol.} = \frac{1}{\pi}\( \({\rm arg}\, \tau_0 - {\rm arg}\, \tau_1\) 
\( \infty\) - \( {\rm arg}\, \tau_0 - {\rm arg}\, \tau_1\)
\( -\infty\)\)$. Again, using the prescription that the argument of $\tau_i$
varies from $-\pi$ to $\pi$, one gets that
\be
Q_{\rm topol.} = 0 \qquad \qquad \mbox{\rm for 
$\(\eps_1\eps_2=1, \bareps_1\bareps_2=-1\)$ or 
$\(\eps_1\eps_2=-1, \bareps_1\bareps_2=1\)$}
\ee
Therefore, such solution corresponds to the scattering of a soliton and a
anti-soliton.

\subsection{The breather solutions}

The space-time dependence of the solutions \rf{soltau0}-\rf{soltault} are
given through the exponentials $\exp \( \gamma_l\( x - v_l t\)\)$, $l=1,2$. 
Therefore,
in order to have solutions periodic in time one needs $\gamma_l  v_l$ to be
pure imaginary. Writing $z_l = \mid z_l\mid e^{i \theta_l}$, one observes from
\rf{gammaandv} that for that to happen one needs $\mid z_l\mid = 1$, and so 
$\gamma_l  v_l = -i m_{\psi} \sin \theta_l$. In order for the solution to
present just one frequency one needs to have 
$\sin \theta_2 = \pm \sin \theta_1$. We do not want $z_1=z_2$ because  
\rf{soltau0}-\rf{soltault} becomes a one-soliton solution. In addition, we do
not want  $z_1=-z_2$ because the solution \rf{soltau0}-\rf{soltault} will
present singularities. Therefore, the only possibilities of having solutions 
periodic in time is to have 
\be
z_1 = e^{i\theta} \; ; \qquad  z_2 = \epsilon \, e^{-i\theta} \; ; \qquad   
\epsilon = \pm 1 
\lab{period1}
\ee
Then one gets from \rf{gammadef} and \rf{gammaandv} that 
\be
\Gamma \( z_1 \) = \gamma x + i \omega t \; \qquad \qquad 
\Gamma \( z_2 \) = \epsilon \( \gamma x - i \omega t \) 
\lab{period2}
\ee
with 
\be
\gamma \equiv m_{\psi} \cos \theta \; \qquad \qquad 
\omega \equiv m_{\psi} \sin \theta 
\lab{period3}
\ee

\subsubsection{The case $\epsilon = 1$}

We are interested in the cases where the field $\vp$ is real, and so from
\rf{tauphinu} we  need $\mid \tau_0 \mid = \mid \tau_1 \mid $. Imposing the
conditions \rf{period1}-\rf{period3} into \rf{soltau0}-\rf{soltau1}
one observes that in order for $\vp$ to be real one needs 
\be
a_{+}^{(2)} a_{-}^{(2)} = \( a_{+}^{(1)} a_{-}^{(1)}\)^* 
\lab{percondreal1}
\ee
and 
\be
\theta + \xi_{+}^{(1)} + \xi_{-}^{(2)} = n \pi \qquad \qquad n \in \IZ 
\lab{percondreal2}
\ee
where we have denoted
\be
a_{\pm}^{(i)} \equiv \mid a_{\pm}^{(i)}  \mid \, e^{i \xi_{\pm}^{(i)}} 
\qquad \qquad i=1,2 
\lab{apmdef}
\ee

It follows that the conditions \rf{percondreal1} and \rf{percondreal2} are
sufficient to make $\tau_1 = \tau_0^*$, and from  \rf{soltau0} one gets
\br
\tau_0 &=& 1 + \( -1\)^{n+1} \, S\,   \( R - \frac{1}{R}\) \, 
\frac{\sin \theta}{\cos^2 \theta} \, e^{2 x m_{\psi} \cos \theta} 
- S^2  \, \tan^4 \theta \,   e^{4 x m_{\psi} \cos \theta} \nonu \\
&+& i \, S\, 
\( \frac{\( -1\)^{n+1}}{\cos \theta } \, 
\( R + \frac{1}{R} \)
- 2 \cos\( 2 \omega t + \xi_{+}^{(1)} + \xi_{-}^{(1)}\) 
 \) e^{2 x m_{\psi} \cos \theta} 
\er
where the integer $n$ is the same as in \rf{percondreal2}, and we have
introduced 
\be
S\equiv \frac{1}{4}\, \mid  a_{+}^{(1)}\mid \mid  a_{-}^{(1)}\mid  
\; ; \qquad \qquad 
R \equiv \frac{\mid a_{+}^{(1)}\mid }{\mid a_{+}^{(2)}\mid } 
\ee

Therefore, using \rf{topological} and \rf{tauphinu} one gets that the
topological charge of such breather solution is 
\be
Q_{\rm topol.} = 2  \( -1\)^{n+1}
\ee
with $n$ given by \rf{percondreal2}. 

Now, if besides the reality of $\vp$, one imposes the condition
\rf{psitrealcond} on the spinors, one gets that the parameters have to satisfy
\be
a_{-}^{(1)} = -  e_{\psi} \, {a_{+}^{(2)}}^* \, e^{i \theta}\; ; \qquad \qquad 
a_{-}^{(2)} = -  e_{\psi} \, {a_{+}^{(1)}}^* \, e^{-i \theta}
\ee
which to be compatible with \rf{percondreal1} and \rf{percondreal2}, one needs 
$\exp \(i\(\theta + \xi_{+}^{(1)} + \xi_{-}^{(2)}\)\) = 
- {\rm sign} \, e_{\psi}$. 
Then, the topological charge becomes 
\be
Q_{\rm topol.} = 2  \, \, {\rm sign} \, e_{\psi}
\ee

Consequently, we have for the breather solutions an effect similar to what
happens in the two-soliton solutions (see discussions below
\rf{onesoltopovppsireal} and \rf{epsiconstr}). For $\vp$ real we have
breathers with topological charges $\pm 2$. However, if one imposes the
reality condition \rf{psitrealcond} on the spinors, one gets that only one
charge is allowed. 

\subsubsection{The case $\epsilon = -1$} 

One can check that for the choice $\epsilon = -1$ in \rf{period1} one can not
obtain non trivial solutions such that $\vp$ is real. However, one does have
that $\vp$ is asymptotically real. It is quite easy to verify that the
topological charge vanishes for any choice of the parameters
$a_{\pm}^{(i)}$. So, 
\be
Q_{\rm topol.} = 0 \; ; \qquad \qquad \mbox{\rm for $\epsilon = -1$, and 
any $a_{\pm}^{(i)}$, $i=1,2$}
\ee

\subsection{The time delays}

Solitons are classical solutions that travel with constant speed without
dispersion and that keep their form under scattering, the effect of it being
only a phase shift or a displacement in its position. We now show that the
solitons  we have been working with are true solitons and indeed fulfill the
above requirements.  We shall calculate the so-called time delays of the
scattering of two solitons, using the procedures of \ct{fring}. 

We consider two solitons that in the distant past are well apart, then colide
near $t=0$ and then separate again in the distant future. Therefore, except
for the region where the scattering occurs, the solitons are free and so
travel with constant velocity. Let us denote the trajectories of one of the
solitons before and after the collision as (since the velocity is the same)
\be
x = vt + x( I) \qquad {\rm and} \qquad x = vt + x( F)
\ee
The lateral displacement at fixed time is measured by
\be
\Delta (x) \equiv x( F) - x( I) 
\lab{latdisp}
\ee
The time delay is defined by 
\be
\Delta(t) \equiv t(F) -t(I) = -\frac{\Delta (x)}{v} 
\lab{timedelay}
\ee
with the intercepts of the trajectories with the time axis being given by 
$t(F)= - x(F)/v$ and $t(I)=-x(I)/v$. The lateral displacement and the time
delay are not Lorentz invariant, and so we consider the invariant 
\be
E \Delta (x) = - p \Delta(t)
\lab{invdelta}
\ee
with $E$ and $p=vE$ being the energy and momentum of the soliton
respectively. Since $E$ is 
positive it follows $\Delta (x)$ has the same sign in any reference frame,
with only its strength being frame dependent. The time delay on the other hand
may change the sign under Lorentz transformations. One can show that the
lateral displacements and time delays for the two solitons participating in
the collision have to satisfy \ct{fring} 
\be
E_1 \Delta_1 (x) + E_2 \Delta_2 (x) = 0 \; ; \qquad \qquad  
p_1 \Delta_1(t) + p_2 \Delta_2(t) =0 
\lab{invdelay}
\ee
with the sub-indices $i$ labeling the quantities associated to particle
$i$. Notice therefore that, since the energies are positive, the lateral
displacements have opposite signs.  
Clearly, in the center of momentum frame (comf), where $p_1 +p_2=0$, 
the time delays are equal, i.e. 
$\Delta_1^{\rm (comf)} (t)=\Delta_2^{\rm (comf)}(t)$. Therefore, from
\rf{invdelta} we have that  $E_1^{\rm (comf)} \Delta_1^{\rm (comf)}(x)/p_1 = 
- E_2^{\rm (comf)} \Delta_2^{\rm (comf)}(x)/p_1 = -\Delta_1^{\rm (comf)}(t) 
=-\Delta_2^{\rm (comf)}(t)$. Consequently, since $\Delta (x)$ has the same
sign in any frame,  it follows that, if particle $1$ moves to the right faster
then particle $2$ (such that $p_1$ is positive in the cmof), then $(-\Delta_1
(x))$,  
$\Delta_2(x)$, $\Delta_1^{\rm (comf)}(t)$ and $\Delta_2^{\rm (comf)}(t)$ all
have the same sign. The physical interpretation of that sign is related to the
character (attractive or repulsive) of the interaction forces. Indeed, if the
force is attractive then particle $1$ will accelerate as it approaches
particle $2$ and then decelerate. That means that $\Delta_1 (x)$ is positive
and so the common sign negative. Therefore, attractive forces lead to a
negative time delay in the center of momentum reference frame, and clearly
repulsive forces a positive time delay. These considerations assume that the
two particles pass through each other, and there is no reflection. However,
when the masses of the two particles are equal there is the possibility of
occurring reflection. 

Let us now evaluate the time delays for the two-soliton solutions given in
\rf{soltau0}-\rf{soltault}. We choose the particle $1$ to move faster to the
right than particle $2$, i.e. $v_1 > v_2$, with $v_1>0$, and therefore from 
\rf{gammaandv} and \rf{zthetarel} $\mid z_1\mid < \mid z_2\mid$ or $\theta_1 <
\theta_2$. We then track the soliton 
$1$ in time \ct{fring}, i.e. we hold $x-v_1 t$ fixed as time varies. Then one
gets $x-v_2t= {\rm const.} + (v_1 - v_2)t$, with  ${\rm const.}=x-v_1 t$.  We
then get that, if $\eps_2 =1$,  $e^{\Gamma_2 (z_2)} \ra 0$ as $t\ra -\infty$,
and $e^{\Gamma_2 (z_2)} \ra \infty$ as $t\ra \infty$. For the case $\eps_2 =
-1$ the limits get interchanged. 

Therefore, taking $\eps_2 =1$, one can check
that the solution \rf{soltau0}-\rf{soltault} 
becomes, in the limit $t\ra -\infty$, the one-soliton solution \rf{tauonesol}
with the identifications 
$a_{\pm}^{(1)} \equiv a_{\pm}$ and $z_1 \equiv z$. Now, in the limit $t\ra
\infty$, one can also verify that the ratios $\tau_0/\tau_1$, $\tau_R/\tau_0$,
$\tau_L/\tau_1$, ${\widetilde \tau}_R/\tau_1$ and ${\widetilde \tau}_L/\tau_0$
for the solutions \rf{soltau0}-\rf{soltault} tend to the corresponding ratios
for the one-soltion solution \rf{tauonesol} with the same identification of
parameters, and with the replacement
\be
e^{\Gamma_1 (z_1)} \ra \( \frac{z_1-z_2}{z_1+z_2}\)^2\,e^{\Gamma_1 (z_1)}
\lab{shiftg1a}
\ee
The only difference being the fact that the ratio $\tau_0/\tau_1$ gets a
relative minus sign, meaning that the $\vp$ field gets shifted by $\pi$ (see
\rf{tauphinu}) during the scattering process. Then one observes from
\rf{tauphinu}-\rf{taupsi} that the relevant effect of the scattering on the
solutions for the fields $\vp$, $\psi$ and ${\widetilde \psi}$ is a lateral
displacement of the soliton $1$ given by 
\be
\gamma_1 \( x - v_1 t\) \ra  \gamma_1 \( x - v_1 t 
+ \frac{1}{\gamma_1} \ln \( \frac{z_1-z_2}{z_1+z_2}\)^2 \) 
\lab{shiftg1b}
\ee
If one takes $\eps_2 =-1$ instead, one observes that the direction  of the
arrow in 
\rf{shiftg1b} reverses. Therefore, using \rf{latdisp}, \rf{gammaandv} and
\rf{zthetarel} one sees that the lateral displacement for the soliton $1$ 
is given by
\be
\Delta_1 (x) = - \frac{\eps_1 \eps_2 }{m_{\psi}\cosh \theta_1}\,  
\ln \(  \frac{e^{\( \theta_1-\theta_2\) /2} 
- \eps_1 \eps_2 e^{-\( \theta_1-\theta_2\) /2 }}
{e^{\( \theta_1-\theta_2\) /2} 
+ \eps_1 \eps_2 e^{-\( \theta_1-\theta_2\) /2 }}\)^2 
\ee
Since $\eps_1 \eps_2=\pm 1$, one observes that $\Delta_1 (x)$ is in fact
independent of such signs, and so
\be
\Delta_1 (x) = - \frac{1 }{m_{\psi}\cosh \theta_1}\,  
\ln \( \tanh\( \frac{\theta_1-\theta_2}{2}\)\)^2 
\lab{deltax1}
\ee
Notice that the solutions \rf{soltau0}-\rf{soltault} are symmetric under the
interchange of the indices $1$ and $2$ of the parameters $a_{\pm}^{(i)}$ and
$z_i$. Therefore, if we track the soliton $2$, i.e. keep $x-v_2t$ fixed as
time varies, but under the same kinematical conditions, i.e. $v_1 > v_2$, with
$v_1>0$, then we obtain that 
\be
\Delta_2 (x) =  \frac{1 }{m_{\psi}\cosh \theta_2}\,  
\ln \( \tanh\( \frac{\theta_1-\theta_2}{2}\)\)^2 
\lab{deltax2}
\ee
If we reverse the kinematical conditions, i.e. take $v_2 > v_1$, with
$v_2>0$, then the signs of both $\Delta_i (x)$, $i=1,2$, reverse. 
The mass of the one-soliton solutions is given in \rf{solmass}, and therefore
the energies of the solitons are $E_i = 2\, k\, m_{\psi}\, \cosh \theta_i$,
$i=1,2$. Consequently, one sees that the $\Delta_i (x)$'s do indeed satisfy
\rf{invdelay}, i.e.
\be
E_1 \Delta_1 (x) = - E_2 \Delta_2 (x) = 
- {\rm sign}\, \( v_1 - v_2\)\, 2\, k\,  
\ln \( \tanh\( \frac{\theta_1-\theta_2}{2}\)\)^2 
\lab{deltaxinv}
\ee
In addition, using \rf{timedelay}, \rf{deltax1} and
\rf{deltax2} one gets that the time delays are given by (assuming $v_1 > v_2$,
with $v_1>0$) 
\be
\Delta_1 (t) = - \frac{1}{m_{\psi}\sinh \theta_1}\,  
\ln \( \tanh\( \frac{\theta_1-\theta_2}{2}\)\)^2 \, ;\quad 
\Delta_2 (t) =  \frac{1}{m_{\psi}\sinh \theta_2}\,  
\ln \( \tanh\( \frac{\theta_1-\theta_2}{2}\)\)^2 
\lab{deltat}
\ee
Notice that the hyperbolic tangent varies from $-1$ to $1$ and therefore the
logarithm of its square is always negative, and so from \rf{deltax1} one sees
that $\Delta_1(x)$ for 
$v_1 > v_2$, with $v_1>0$, is positive.   
Therefore,  from  the considerations made above we conclude
that the forces between the solitons is attractive. In addition, it is
independent of their topological charges. In fact, the time delays we have
obtained coincide with those of the sine-Gordon theory by identifying $k$ with
a positive constant multiplied by the inverse of the square of the sine-Gordon
coupling constant.

\appendix

\section{Appendix: The zero curvature}
\label{appa}

The equations of motion of the theory \rf{lagrangian} can be represented as a
zero curvature \rf{zc} with connections given by \ct{matter}
\br
A_{+} = - B\, \( E_2 + F^{+}_1 \)  \, B^{-1} \, , \qquad
A_{-} = - \pa_{-} B \,  B^{-1} + E_{-2} + F^{-}_1.
\lab{gp}
\er
where 
\be
B= e^{i\vp H^0} \, e^{\( \nu -{i\o 2} \vp \) C} \, e^{\eta Q} \qquad \qquad 
E_{\pm 2} \equiv {1\o 4}m_{\psi} \, H^{\pm 1} 
\lab{e2}
\ee 
and 
\be
F^{+}_1 = \sqrt{i m_{\psi}}\( \psi_R\, E_+^0 +
\widetilde \psi_R E_-^1\)\, ,\quad
F^{-}_1 = \sqrt{i m_{\psi}}\( \psi_L\, E_+^{-1} -
\widetilde \psi_L\, E_-^0 \) ,
\ee
We have written  the Dirac spinors as 
\br
\psi = \(
\begin{array}{c}
\psi_R\\
\psi_L
\end{array}\) \, ; \qquad
\widetilde \psi = \(
\begin{array}{c}
\widetilde \psi_R\\
\widetilde \psi_L
\end{array}\)
\er
and have denoted by $H^n$, $E_{\pm}^n$, $D$ and $C$ the Chevalley
basis generators of the $sl(2)$ affine Kac-Moody algebra. The
commutation relations are 
\br
\lb H^m \, , \, H^n \rb &=& 2 \, m \, C \, \d_{m+n,0},
\lab{sl2a}\\
\lb H^m \, , \, E^n_{\pm} \rb &=& \pm 2 \, E^{m+n}_{\pm},
\lab{sl2b}\\
\lb E^m_{+} \, , \, E^n_{-} \rb &=& H^{m+n} + m \, C \, \d_{m+n,0},
\lab{sl2c}\\
\lb D \, , \, T^m \rb &=& m \, T^m \, , \qquad T^m \equiv H^m , E_{\pm}^m;
\lab{sl2d}
\er
The generator $Q$ is the grading operator for the principal gradation and
given  by $Q\equiv \h H^0 + 2 D$. 
 
\vspace{1cm}

\noindent {\bf Acknowledgements}

H.S.B.A. is supported by a Fapesp grant, and L.A.F. is partially supported by
CNPq. The authors are grateful to J.F. Gomes, M.A.C. Kneipp, D.I. Olive,
J. S\'anchez Guill\'en, G. Sotkov and A.H. Zimerman for many helpful
discussions.

\end{document}